\documentclass[]{aa}  
\usepackage{amsmath}
\usepackage[normalem]{ulem}

\usepackage[dvipsnames]{xcolor}
\usepackage{float}
\usepackage{graphicx}

\usepackage{lscape}

\usepackage{graphicx}
\usepackage{multicol, blindtext}
\usepackage{breqn}
\usepackage{txfonts}
\usepackage[]{hyperref}
\def\Bvec{$\boldsymbol{B}$}

\begin{document}

    \title{Stratification of canopy magnetic fields in a plage region}
    \subtitle{Constraints from a spatially-regularized weak-field approximation method}

   \author{
          Roberta Morosin
          \and
          Jaime de la Cruz Rodr\'iguez
          \and
          Gregal J.~M.~Vissers
          \and
          Rahul Yadav
          }

   \institute{Institute for Solar Physics, Dept. of Astronomy, Stockholm University,               AlbaNova University Centre, SE-10691 Stockholm, Sweden\\
                \email{roberta.morosin@astro.su.se}
             }

   \date{Received XXX; accepted XXX}

 
  \abstract
 {The role of magnetic fields in the chromospheric heating problem remains greatly unconstrained. Most theoretical predictions from numerical models rely on a magnetic configuration, field strength, and connectivity; the details of which have not been well established with observational studies for many chromospheric scenarios. High-resolution studies of chromospheric magnetic fields in plage are very scarce or non existent in general.}
 {Our aim is to study the stratification of the magnetic field vector in plage regions. Previous studies predict the presence of a magnetic canopy in the chromosphere that has not yet been studied with full-Stokes observations. We use high-spatial resolution full-Stokes observations acquired with the CRisp Imaging Spectro-Polarimeter (CRISP) at the Swedish 1-m Solar Telescope in the \ion{Mg}{i}~5173~\AA,  \ion{Na}{i}~5896~\AA\, and \ion{Ca}{ii}~8542~\AA\ lines.}
 {We have developed a spatially-regularized weak-field approximation (WFA) method, based on the idea of spatial regularization. This method allows for a fast computation of magnetic field maps for an extended field of view. The fidelity of this new technique has been assessed using a snapshot from a realistic 3D magnetohydrodynamics simulation.}
 {We have derived the depth-stratification of the line-of-sight component of the magnetic field from the photosphere to the chromosphere in a plage region. The magnetic fields are concentrated in the intergranular lanes in the photosphere and expand horizontally toward the chromosphere, filling all the space and forming a canopy. Our results suggest that the lower boundary of this canopy must be located around $400-600$\,km from the photosphere. The mean canopy total magnetic field strength in the lower chromosphere ($z\approx760$~km) is 658\,G. At $z=1160$~km, we estimate $<B_\parallel>\approx 417$~G.}
 {In this study we propose a modification to the WFA that improves its applicability to data with a worse signal-to-noise ratio. We have used this technique to study the magnetic properties of the hot chromospheric canopy that is observed in plage regions. The methods described in this paper provide a quick and reliable way of studying multi layer magnetic field observations without the many difficulties inherent to other inversion methods.}

   \maketitle
%

\section{Introduction}

The concept of plage in solar physics was coined more than 100 years ago from the analysis of low spatial-resolution $H\alpha$ observations. These regions showed enhanced chromospheric line emission in many diagnostics such as the $H\alpha$ line \citep{deslandres1893}. Our understanding of the physical processes that are at work in plage is still not complete. For instance, the origin of the large values of microturbulence that are required to model chromospheric observations remains unclear \citep{1974SoPh...39...49S,2015ApJ...809L..30C}.

In the photosphere, reconstructions of the magnetic field vector indicate a strongly vertical orientation. On the other hand, the balance between magnetic pressure and gas pressure in the chromosphere is dominated by the former and the magnetic field is expected to expand horizontally, forming a canopy above the photosphere where the magnetic concentrations are confined in the intergranular lanes. The chromospheric canopy is hot and it leaves strong imprints in many chromospheric lines \citep{1974SoPh...39...49S,1991A&A...250..220S,2013ApJ...764L..11D,2015ApJ...809L..30C}. The first direct observational evidence of the expansion of the magnetic field was reported by \citet{1994ApJ...424.1014S}. More recent studies with high spatial resolution observations have reconstructed this expansion in the photosphere, where the reconstructed magnetic field concentrations become slightly more space-filling in the upper photosphere \citep{2015A&A...576A..27B}.

The chromospheres of plage have been of particular interest for researchers because strong radiative cooling is predicted through the \ion{Ca}{ii}~H\&K lines, the \ion{Ca}{ii} infrared triplet lines, and the \ion{Mg}{ii}~h\&k lines \citep{1981ApJS...45..635V,2012A&A...539A..39C} and that energy must be provided efficiently by an energy transport mechanism. Physical processes related to the presence of strong magnetic fields
are obvious candidates to deliver this energy \citep[see, e.g.,][]{2008ApJ...680.1542H,2010ApJ...720..776J,2011ApJ...736....3V,2018ApJ...862L..24P}. Therefore, understanding the magnetic field topology and its connectivity between the photosphere and the chromosphere could also help to understand the energy balance in these regions. Interestingly, the gradient of the magnetic field is large at the lower edge of the magnetic canopy and, therefore, a large electric current density is predicted by Ohm's law ($\boldsymbol{j} = \nabla\times\boldsymbol{B}/\mu_0$). The presence of those electric currents can lead to energy dissipation at the lower boundary of the chromosphere. The methods presented in this study can greatly contribute to understanding the energetic contribution of these currents to the chromospheric energy budget, which is our ultimate goal.

Although inversion methods have allowed retrieving plage magnetic fields in the photosphere in great detail, in the chromosphere the situation has been different. The limited selection of spectral lines with sufficient magnetic sensitivity combined with complex non local/non equilibrium physical processes that must be accounted for in order to model most of these lines have hindered enormously this line of research \citep[see, e.g.,][and references therein]{2017SSRv..210..109D,2019ARA&A..57..189C}. Inversion codes that allow including non local thermodynamical equilibrium effects (NLTE) exist in different implementations but in all cases they require a large amount of computational resources in order to process a large field-of-view \citep{2015A&A...577A...7S,2018A&A...617A..24M,delacruz2019}. Another limitation of these codes is that they assume a smooth depth-stratified atmosphere that might not always be able to describe complex stratifications in all physical parameters, and that can influence their ability to properly reconstruct the magnetic field.

The weak-field approximation \citep[WFA;][]{landi1973} provides an alternative method to quickly obtain magnetic field estimates whenever certain assumptions are satisfied (see \S\ref{sec:wfa}). The WFA does not require knowledge of atom level populations to derive the emerging intensities in Stokes $Q$, $U$ and $V$. This method has been used extensively to analyze observations in the \ion{Ca}{ii}~8542~\AA\ and the \ion{Mg}{i}~b$_1$ lines \citep[e.g.,][]{pietarila2007,2009ApJ...700.1391M,kleint2017,2018A&A...609A..14R,centeno}. However, the WFA is equally affected by low signal-to-noise in observations as inversions are. 

A new generation of inversion codes have made use of the idea of spatial coherency to improve the fidelity of their algorithms. A study by \citet{2012A&A...548A...5V} proposed to couple the solution of neighboring pixels using the telescope point-spread-function. A different approach has also been suggested by \citet{2015A&A...577A.140A}, using transformations to a different space where their model parameters are sparse and a decreased number of free parameters could be used to fit their data. In a recent paper, \citet{2019A&A...631A.153D} studied the effect of Tikhonov regularization in the inversion of spatially resolved spectra using a non linear Milne-Eddington model atmosphere in combination with a Levenberg-Marquardt algorithm. They showed that the quality of the inferred model parameters is greatly improved when spatial constraints are present in the fitting algorithm. 

In this paper, we propose to use a spatially-regularized WFA algorithm, based on the imposition of Tikhonov regularization \citep{Tikhonov77} to infer the magnetic field vector from spectropolarimetric (chromospheric) observations. The underlying idea is to find magnetic field values that are not only compatible with the spectrum of the local pixel, but that also keep some spatial coherence with the nearest neighboring pixels. The latter is achieved by penalizing strong horizontal gradients in the reconstructed magnetic field. Random noise solution with large amplitudes are therefore not compatible with these spatial constraints. Here we show that this modification of the WFA least-squares equations allows improving the fidelity of the reconstruction, while also extending the usability of this technique to data with a lower signal-to-noise ratio. 

We have used this new method with high spatial resolution observations acquired in the \ion{Mg}{i}~5173~\AA, the \ion{Na}{i}~5896~\AA\ and the \ion{Ca}{ii}~8542~\AA\ lines. The combination of these lines allows sampling the solar atmosphere in the photosphere, the upper photosphere and the lower chromosphere with some redundancy and depth overlap. We study the expansion of the canopy as a function of height and reconstruct a vertical slice of the line-of-sight magnetic field stratification.

\section{Spatially-regularized weak-field approximation}\label{sec:wfa}
    \begin{figure}
    \centering
    \includegraphics[width=\columnwidth]{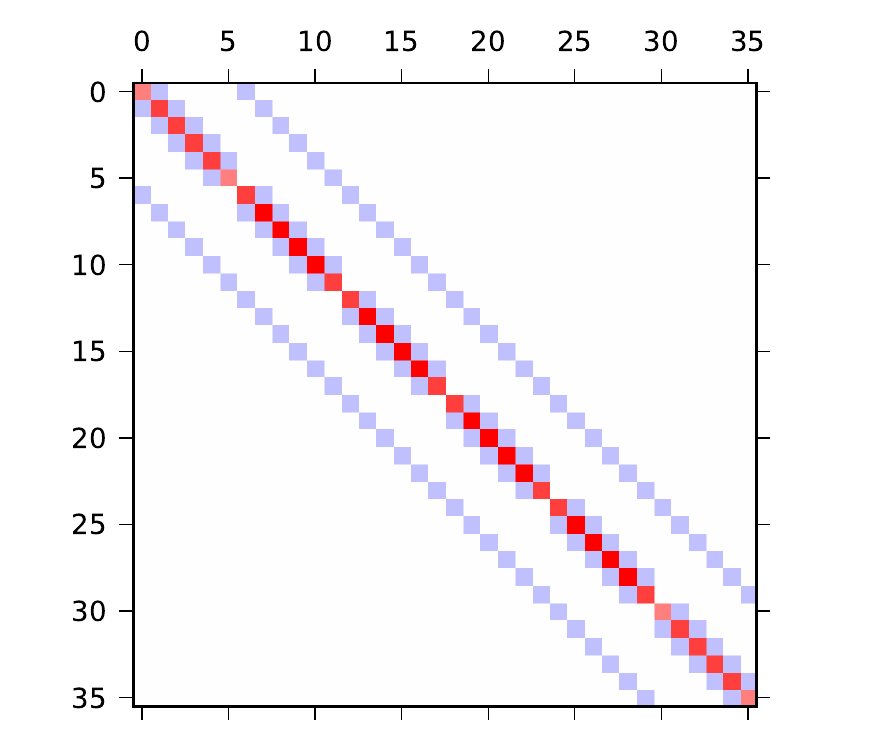} 
        \caption{Regularization terms of the left-hand-side matrix for a FOV of $6\times6$ pixels. Each row of the matrix represents how one pixel is coupled to its neighbors. The image is scaled between $\pm 4\alpha$. Red elements are positive and blue elements are negative.}
              
         \label{Amatrix}
    \end{figure}
       
   The weak-field approximation is an approximate solution of the radiative transfer equation that allows predicting Stokes~$Q$, $U$ and $V$ as a function of Stokes~$I$ and its derivatives as a function of wavelength \citep{landi1973}. The assumption underlying the WFA is that the magnetic field does not change with depth and that the splitting induced by the Zeeman effect is significantly smaller than the Doppler width of the line. In strong chromospheric lines, like the \ion{Ca}{ii}~8542~\AA\ line, not having a constant magnetic field with depth would mean that the wings, sensitive to the photosphere, would be affected by a different value of the magnetic field than the core of the line, which is sensitive to the chromosphere. The latter condition, ($\Delta \lambda_B \ll \Delta \lambda_D $) defines very well the regime of applicability of the approximation \citep{asensioramos}:
    \begin{equation}
            B < \frac{4\pi m_e c}{\bar{g} \lambda_0 e_0}
            \sqrt{\frac{2kT}{M}+ v^2_{mic}}
             \,,
    \end{equation}
   where $\bar{g}$ is the effective Land\'{e} factor (that in the case of 8542 \r{A} is equal to 1.10), $\lambda_0$ is the central wavelength of the line, $m_e$ and $e_0$ are respectively the electron mass and charge, $c$ is the speed of light, $k$ is the Boltzmann constant, $T$ is the temperature, $M$ is the mass of the atomic element corresponding to the line and $v_{mic}$ the microturbulent velocity. 

   In strong magnetic concentrations like sunspots, photospheric lines frequently exceed this limit \citep[e.g.,][]{2013A&A...557A..24V}, but in the chromosphere the magnetic field is typically weaker \citep[e.g.,][]{2016A&A...596A...8J}. In addition, the spectral lines that sample chromospheric conditions normally have a larger Doppler width than photospheric ones as well as relatively low Land\'e factors. So, the WFA is well-suited for chromospheric lines in the Zeeman regime, for example \ion{Ca}{ii}~8542~\AA. In fact, the WFA has been employed regularly to extract the magnetic field from observations in this line \citep[e.g.,][]{pietarila2007,jaime2013,centeno,kleint2017,kuridze2019}. 
   
   For Stokes $Q$ and $U$ an additional constraint is imposed and the approximation only works if there are no (strong) velocity gradients in the atmosphere, which can lead to opacity jumps that break the validity of the approximation.
   
   In this study we have followed the derivation presented in \citet{landi} to obtain the differential equations that associate the different Stokes parameters ($Q$, $U$, $V$) to the derivative of Stokes $I$ with respect to wavelength. In these equations, the two components of the magnetic field appear: the line-of-sight component connects Stokes $V$ to the derivative of Stokes $I$ (section \ref{longitudinal}), while the horizontal one connects Stokes $Q$ and $U$ to the derivative of Stokes $I$ (section \ref{horizontal}). We have also calculated the azimuth of \Bvec\ in the plane of the sky (section \ref{azimuth}). A similar derivation, along with uncertainties estimates, is provided in \citet{martinez}.
   
   Here, we have derived our mathematical expressions imposing spatial constraints from the nearest-neighbor pixels by minimizing the difference between their values with a certain weight. The latter is done in the form of spatial Tikhonov regularization \citep{Tikhonov77}.
   In principle, given that our observations are critically sampled spatially and the magnetic field is expected to be relatively smooth in the chromosphere, the inferred magnetic field cannot be extremely different between pixels that are close to each other. In particular we have considered the four nearest neighbors: the pixel on the right, the left, above and below. The value of the magnetic field (longitudinal and transverse) in a certain pixel is then also determined by the solution of the surrounding pixels and it cannot be computed individually anymore, like in the case of the standard WFA. As a result, the problem becomes global and the magnetic field must be solved for the entire field of view simultaneously. Taking into account the values of nearby pixels is expected to be a decisive improvement of the fidelity of the WFA solution as it should limit the impact of noise.
 
    \begin{figure*}[!ht]
    \centering
    \includegraphics[width=\textwidth]{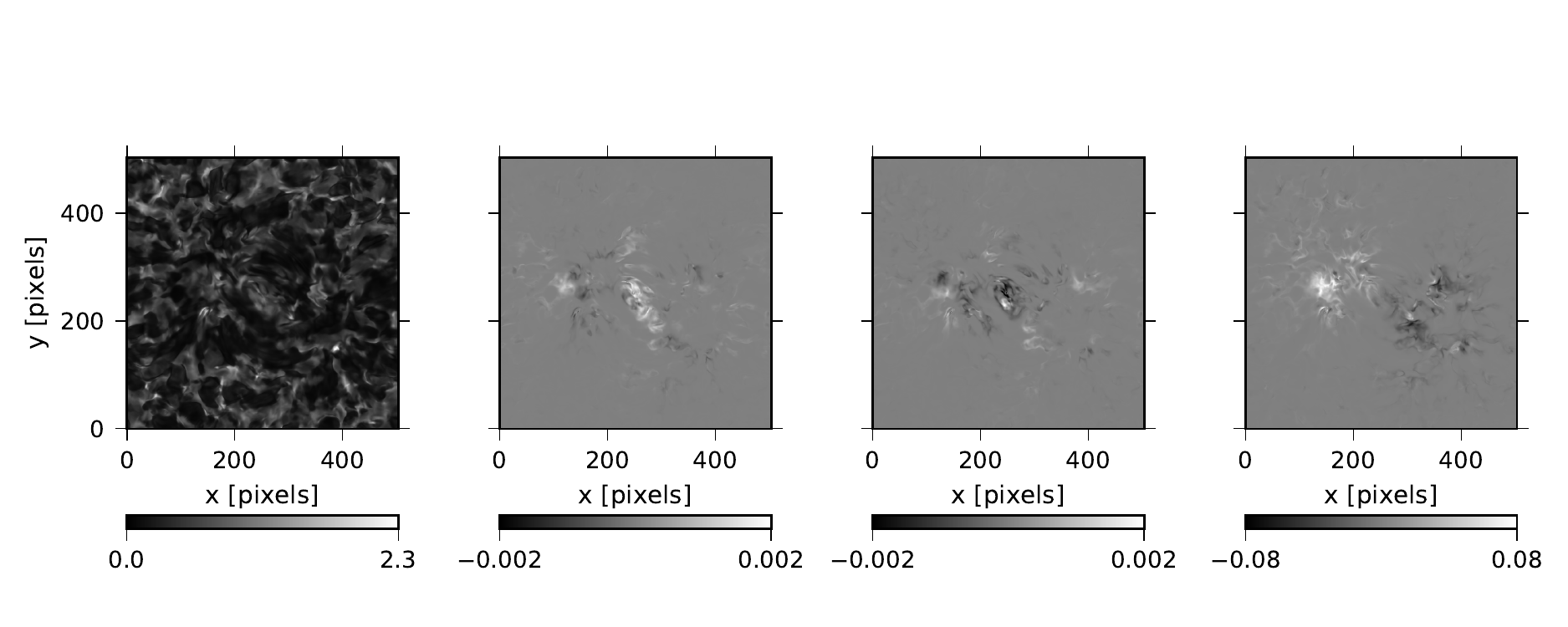} 
        \caption{Synthetic \ion{Ca}{ii}\,\,8542\,\AA\ Stokes components at line center from the Bifrost snapshot. {\it{Left to right:\/}} Stokes $I, Q, V$ and $U$.}
         \label{syn_stokes}
    \end{figure*}
    
\subsection{Line-of-sight component}\label{longitudinal}
    In this section, we focus on the line of sight, or longitudinal, component of \Bvec.
    The differential equation of the first order gives the relation between Stokes $V$ and the derivative of Stokes $I$:
    \begin{equation}\label{vcomp}
        V(\lambda)=-\Delta \lambda_B \bar{g} \cos{\theta} \frac{dI}{d\lambda}
    \end{equation}
    where $\Delta \lambda _B$ is given by:
    \begin{displaymath}
        \Delta \lambda _B = 4.6686 \cdot 10^{-13}\lambda_0 ^2 B 
    \end{displaymath}
    Equation \eqref{vcomp} can be written in a more compact way:
    \begin{equation}
        V(\lambda)=-C_1 B_{\parallel} \frac{dI}{d\lambda}
    \end{equation}
    where the constant $C_1$ includes all the constant terms that appear in Eq.~(\ref{vcomp}) and $B_{||}$ is the line-of-sight component of \Bvec\ in which is included $\cos{\theta}$, with $\theta$ the angle between the observer's line-of-sight and the normal to the solar surface.
    
    {\it A priori\/}, the standard WFA treats each pixel in the image individually, ignoring the rich and redundant information that is present in the spatial dimensions.
    In the standard case, the WFA is applied to observations by defining a merit penalty function and trying to minimize it, for example, for longitudinal component of the magnetic field:
    \begin{equation}
        \chi ^2 = \frac{1}{N}\sum _{i=\lambda_0}^{\lambda_n} \left(\frac{V_{obs}^i-V_{synt}^i}{\sigma_i}\right)^2= \frac{1}{N}\sum _{i=\lambda_0}^{\lambda_n}\frac{1}{\sigma_i^2}\left(V_{obs}^i -\biggl(-C_1 \frac{dI_i}{d\lambda}B_{\parallel}\biggr)\right)^2.\label{eq:wfa_vanilla}
    \end{equation}
    Hereafter, we assume that each data point is divided by the corresponding $\sqrt{N}\sigma_i$, dropping those terms from the notation. Since both Stokes $V$ and $I$ are observed quantities, we can derive the longitudinal component of the magnetic field by simply minimizing the value of $\chi ^2$, which in this case is trivial because the dependence on the model parameters is linear:
    \begin{equation}\label{eq:blos}
        B_{\parallel}= \frac{\sum_{i} V_i \frac{dI_i}{d\lambda}}{C_1\sum_{i}\biggl(\frac{dI_i}{d\lambda}\biggr)^2}.
    \end{equation}
    
    The least-squares formula in Eq.~(\ref{eq:blos}) has been used widely 
    in solar and stellar applications in the past. 
    With the following definition the solution is forced to be compatible also with the surrounding pixels,
    %
   imposing spatial constraints from the nearby pixels by reducing the difference with a certain weight, $\alpha$:
    \begin{multline}\label{chi_long}
        \chi ^{2} = \sum _{i=\lambda_0}^{\lambda_n}\biggl(V_i - C_1 \frac{dI_i}{d\lambda}B_{\parallel}^{(x,y)}\biggr)^2 + 
        \alpha \biggl[ \left(B_{\parallel}^{(x,y)}-B_{\parallel}^{(x,y-1)}\right)^2 \\
        + \left(B_{\parallel}^{(x,y)}-B_{\parallel}^{(x,y+1)}\right)^2 + \left(B_{\parallel}^{(x,y)}-B_{\parallel}^{(x-1,y)}\right )^2 + \left(B_{\parallel}^{(x,y)}-B_{\parallel}^{(x+1,y)}\right)^2 \biggl]
      \end{multline}
    Here we have added two superscripts to $B_{||}$ that correspond to its \emph{(x,y)} location in the field of view.
    
    Taking the derivative of the previous formula with respect to the line-of-sight components of the magnetic field in a specific pixel yields:
    \begin{multline}\label{spatconstraints}
        \biggl[ C_1^2 \sum_i \biggl(\frac{dI_i}{d\lambda} \biggr)^2 + 4 \alpha \biggr] B_{\parallel}^{(x,y)} - \alpha B_{\parallel}^{(x,y-1)} - \alpha B_{\parallel}^{(x,y+1)} - \alpha B_{\parallel}^{(x-1,y)} \\
        -\alpha B_{\parallel}^{(x-1,y)} =C_1 \sum_i V_i \frac{dI_i}{d\lambda}.
    \end{multline}
    This equation refers to the pixel $(x,y)$, so, if the image has dimensions $(n_x, n_y)$, we obtain $(n_x \cdot n_y)$ equations. The factor \emph{4}, in this case, that multiplies $\alpha$ with the number of subtracted terms in the left part of the equation depends on the position of the pixel in the image. Pixels in the corners and on the edges of the field-of-view (FOV) have a different factor and number of terms. It is also clear that every equation is bound to the others through the values of the neighboring pixels, so it is not possible anymore to obtain the longitudinal component of the magnetic field for every pixel separately. Instead, to get the solution of the problem, we have to solve a sparse linear system of equations of the form $\mathbf{A}\boldsymbol{x}=\boldsymbol{b}$.
   
    The sparse matrix \textbf{A} contains all the terms that multiply the values of the magnetic field: the diagonal term that depends on the derivative of Stokes $I$ with respect to the wavelength plus a contribution of the local magnetic field to the regularization term ($4\alpha$). Additionally we get four non diagonal terms that are proportional to $-\alpha$ and that account for the dependence on the solution for neighboring pixels. The matrix $\mathbf{A}$ has dimensions $(n_x \cdot n_y, n_x \cdot n_y)$. Each row of the matrix maps the solution of our model into one observed pixel. In each row, only five elements (or fewer) are non zero, so we can store this matrix in a compact way, discarding all zeros. Figure~\ref{Amatrix} visualizes how the regularization terms in the \textbf{A} matrix are structured for an example field-of-view of ($6\times6$) pixels: the dimensions are (36,36) and the values on the diagonal change according to the number of the other non zero elements. The matrix \textbf{x} includes all the values of the line-of-sight component of the magnetic field, which is the quantity we want to derive. The vector $\boldsymbol{b}$ contains all the right-hand side terms in Eq.~(\ref{spatconstraints}).

    We use the biconjugate gradient stabilized method (BiCGSTAB) \citep{vandervorst} to solve the sparse linear system of equations. This is an iterative method to solve symmetric and non symmetric linear systems and it provides very good convergence properties among other methods used to solve these types of linear systems \citep{Saad2003}.
    
\subsection{Horizontal component}\label{horizontal}
    The perturbation analysis of the polarized radiative transfer equation presented by \citet{landi1973} allows obtaining a second-order solution that expresses Stokes~$Q$ as a function of the derivative of Stokes~$I$. The dependence is different in the wings $(\lambda \gg \Delta\lambda_D)$ and in the core $(\lambda \ll \Delta \lambda_D)$:
    \begin{eqnarray}
        Q(\lambda_0) &=& -\frac{1}{4}\Delta\lambda_B^2 \bar{G} \sin^2{\theta}
        \cos{2\phi_B} \biggl(\frac{d^2I}{d\lambda^2}\biggl) \label{linecent} \\
        Q(\lambda_w) &=& \frac{3}{4}\Delta\lambda_B^2 \bar{G} \sin^2{\theta} \cos{2\phi_B} \frac{1}{\lambda_w-\lambda_0} \biggl(\frac{dI}{d\lambda}\biggl) \label{wingsq}
    \end{eqnarray}
    where $\bar{G}$ evaluates the sensitivity of linear polarization to \Bvec, which just depends on the type of the transition. Equations~(\ref{linecent}) and (\ref{wingsq}) are used in the center and in the wing of the line, respectively. In this study we only use the line wing Eq.~(\ref{wingsq}), since the applicability of  Eq.~(\ref{linecent}) is restricted to the line core, while common observational samplings are likely to cover more wavelength points outside $\Delta\lambda_D$. In both equations there is a term depending on $\phi_B$, which is the azimuth angle of \Bvec\ with respect to a reference direction \citep{martinez}.
    
    Similarly, the line-wing equation for Stokes~$U$ can be defined as:
    \begin{equation}\label{wingsu}
        U(\lambda_w) = \frac{3}{4}\Delta\lambda_B^2 \bar{G} \sin^2{\theta} \sin{2\phi_B} \frac{1}{\lambda_w-\lambda_0} \biggl(\frac{dI}{d\lambda}\biggl).
    \end{equation}

    To simplify we can define two different components of the horizontal magnetic field, one with the contribution of Stokes~$Q$ and the other one with the contribution of Stokes~$U$:
    \begin{eqnarray}
        B_{\bot_{Q}} &=& B \sin{\theta} \cos{\phi_B}\\
        B_{\bot_{U}} &=& B \sin{\theta} \sin{\phi_B}
    \end{eqnarray}
    As in section \ref{longitudinal}, we obtain the traditional WFA for the transverse magnetic field component by defining a merit function and minimizing it. In this case, we have obtained it for both contributions of $Q$ and $U$ to $B_\bot$:
    \begin{eqnarray}
        \left(B^2_{\bot_{Q}}\right)^2 &=& \left(\frac{\sum_i Q_i \biggl(\frac{d\bar{I}_i}{d\lambda}\biggr)}{C_2\sum_i \biggl(\frac{d\bar{I}_i}{d\lambda}\biggr)^2}\right)^2 ,\label{eq:Bq}\\ 
        \left(B^2_{\bot_{U}}\right)^2 &=& \left(\frac{\sum_i U_i \biggl(\frac{d\bar{I}_i}{d\lambda}\biggr)}{C_2\sum_i \biggl(\frac{d\bar{I}_i}{d\lambda}\biggr)^2}\right)^2 ,\label{eq:Bu}
    \end{eqnarray}
    where $C_2$ includes all the constant terms in Eqs.~(\ref{wingsq}) and (\ref{wingsu}) and it is different from $C_1$ due to the different Land\'e factors for Stokes $Q$ and $U$. On the other hand, the term $\biggl(\frac{1}{\lambda_w-\lambda_0}\biggr)$ is now included in the derivatives of the intensity and we have changed the notation to $\bar{I}$ in order to reflect this contribution of the wavelength term. The total horizontal component of the magnetic field then follows as:
    \begin{equation}\label{bhor}
        B^2_{\bot}= \sqrt{\left(B^2_{\bot_{Q}}\right)^2+\left(B^2_{\bot_{U}}\right)^2}
    \end{equation}
    \citep[see also, e.g.,][]{martinez}. In contrast, many studies adopt an alternative formulation that makes use of $L=\sqrt{Q^2 + U^2}$ instead of using separate estimates of $B_\bot$ from $Q$ and $U$ \citep[e.g.,][]{asensioramos,Jennerholm2014,kuridze2018,centeno}. When faced with noisy data this leads to issues because it always predicts a non zero value of $L$ even if the $Q$ and $U$ observations are pure noise. As a result, when applied to very noisy data, the reconstructed $|B_\bot|$ maps predict strong magnetic fields everywhere. In our case, Eqs.~(\ref{eq:Bq}) and (\ref{eq:Bu}) keep the sign of $Q$ and $U$ in the numerator and, in noisy cases, the summation should average to a number close to zero if $Q$ and $U$ are pure noise and the denominator is non zero. 
    
    We can define a modified merit function for the transverse field that accounts for regularization in exactly the same way as for the parallel component of the magnetic field in Section~\ref{longitudinal}, adding a term dependent on the regularization parameter $\alpha$. Like in the case of $B_{\parallel}$, $\alpha$ is a constant that has to be calibrated according to the observational data that are being processed. Since there are two different contributions to $B_{\bot}$, the solutions for Stokes~$Q$ and $U$ need to be calculated separately with two different linear systems of equations: $\mathbf{A}_{Q} \boldsymbol{x}_{Q} = \boldsymbol{b}_{Q}$ and $\mathbf{A}_{U} \boldsymbol{x}_{U} = \boldsymbol{b}_{U}$. Since Stokes $Q$ and $U$ are linearly dependent on $B_\bot^2$, we make a variable change $\bar{B}_\bot = B_\bot^2$ to simplify the notation. The equation per pixel for the Stokes~$Q$ component of $\bar{B}_{\bot}$ is:
    \begin{multline}\label{chi_q}
        \Biggl[ C_2^2 \sum_i \biggl( \frac{d\bar{I}_i}{d\lambda} \biggr)^2 + 4 \alpha \Biggr] \left(\bar{B}^{(x,y)}_{\bot_Q}\right)^2- \alpha \left(\bar{B}_{\bot_Q}^{(x,y-1)}\right)^2 - \alpha \left(\bar{B}_{\bot_Q}^{(x,y+1)}\right)^2 \\- \alpha \left(\bar{B}_{\bot_Q}^{(x-1,y)}\right)^2 
        -\alpha \left(\bar{B}_{\bot_Q}^{(x-1,y)}\right)^2=C_2\sum_i Q_i \frac{d\bar{I}_i}{d\lambda},
    \end{multline}
    while the one for the Stokes U component of $\bar{B}_{\bot}$ is:
    \begin{multline}\label{chi_u}
        \Biggl[ C_2^2 \sum_i \biggl( \frac{d\bar{I}_i}{d\lambda} \biggr)^2 + 4 \alpha \Biggr] \left (\bar{B}^{(x,y)}_{\bot_U}\right)^2- \alpha \left(\bar{B}_{\bot_U}^{(x,y-1)}\right)^2 - \alpha \left(\bar{B}_{\bot_U}^{(x,y+1)} \right)^2 \\ - \alpha \left(\bar{B}_{\bot_U}^{(x-1,y)}\right)^2 
        -\alpha \left(\bar{B}_{\bot_U}^{(x-1,y)}\right)^2=C_2\sum_i U_i \frac{d\bar{I}_i}{d\lambda}.
    \end{multline}
    As mentioned previously the number of terms with $\alpha$ and the factor 4, that multiplies $\alpha$ in square brackets, in the left part of the equations depend on the position of the pixel in the field of view.
    Each component has been calculated separately using the BiCGSTAB method, and combined afterwards using Eq.~(\ref{bhor}).
    
\subsection{Azimuth}\label{azimuth}
    The azimuth has already been introduced in Eqs.~(\ref{linecent}), (\ref{wingsq}) and (\ref{wingsu}). 
    By the definition of $Q$ and $U$, dividing Eq.~(\ref{wingsu}) by (\ref{wingsq}), we obtain:
    \begin{equation}
        \tan{2\phi_B} = \frac{U(\lambda)}{Q(\lambda)}
    \end{equation}
    By replacing the previously derived formulas, we can obtain the standard WFA estimate of the azimuthal angle of the magnetic field:
    \begin{equation}\label{azimuth1}
        \phi_B = \frac{1}{2} \arctan \frac{\sum_i U_i \biggl(\frac{dI_i}{d\lambda}\biggr)}{\sum_i \biggl(\frac{dI_i}{d\lambda}\biggr)^2} \frac{\sum_i \biggl(\frac{dI_i}{d\lambda}\biggr)^2}{\sum_i Q_i \biggl(\frac{dI_i}{d\lambda}\biggr)} = \frac{1}{2}\arctan \frac{\sum_i U_i \biggl(\frac{dI_i}{d\lambda}\biggr)}{\sum_i Q_i \biggl(\frac{dI_i}{d\lambda}\biggr)}.
    \end{equation}
    
    Once we introduce the $\alpha$ parameter and the magnetic field of the nearby pixels it is no longer possible to simplify the term $\sum_i \left(\frac{dI_i}{d\lambda}\right)^2$ of Eq.~(\ref{azimuth1}), because it is precisely through that term that we impose the spatial constraints. Instead, we have to compute $\bar{B}_{\bot_Q}$ and $\bar{B}_{\bot_U}$ individually using Eq.~(\ref{chi_q}) and (\ref{chi_u}) and then calculate the azimuth directly:
    \begin{equation}
        \phi_B = \frac{1}{2}\arctan\frac{{\bar{B}_{\bot_U}}}{{\bar{B}_{\bot_Q}}}.
    \end{equation}
        \begin{figure}    
    \centering
    \includegraphics[width=\columnwidth]{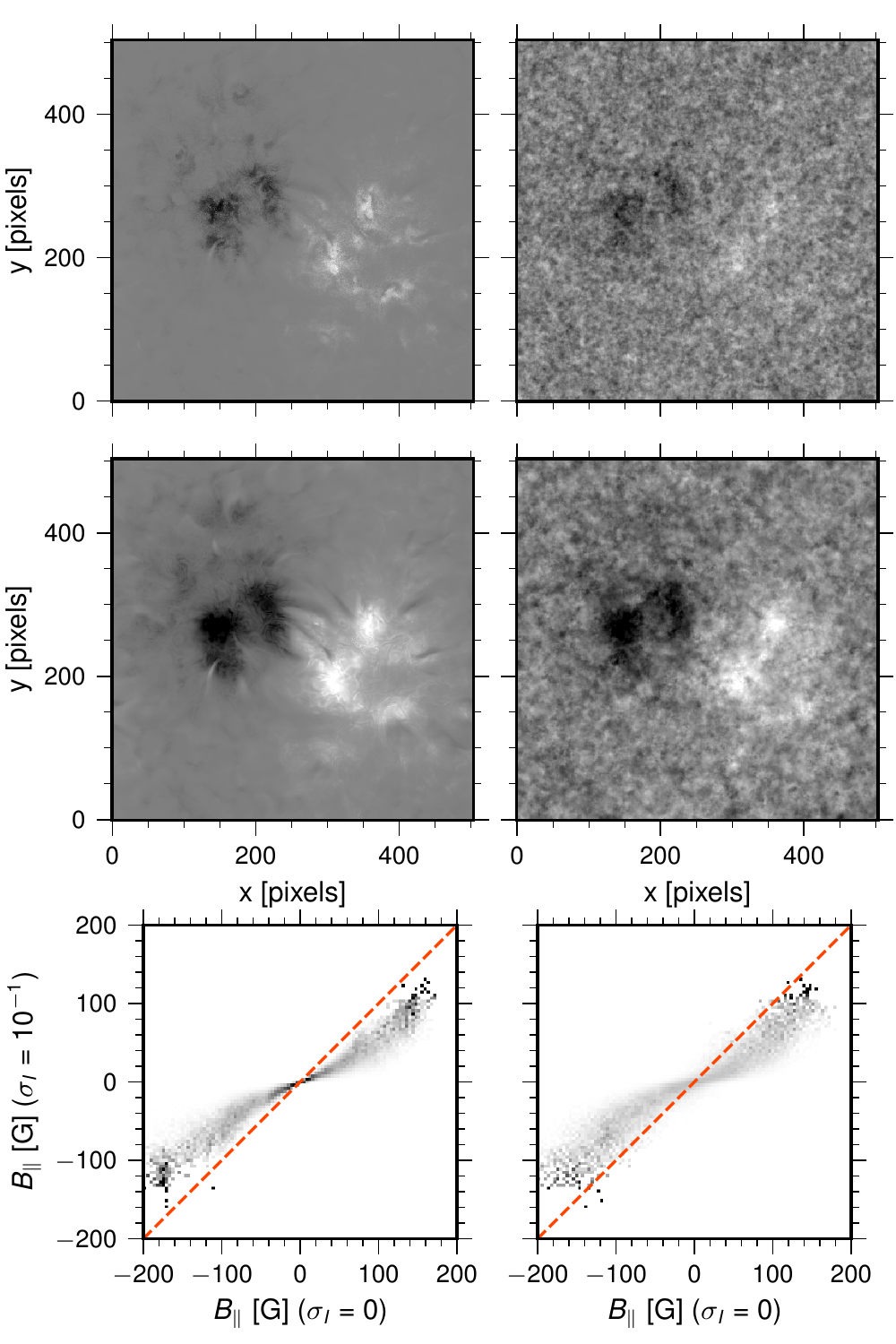}
        \caption{Effect of enhanced noise levels in Stokes $I$.
        {\it Top two rows:\/} Maps of $B_{\parallel}$ assuming $\sigma_I = 10^{-1}$ ({\it first row\/}), $\sigma_I = 0$ ({\it second row\/}),  $\sigma_V = 0$ ({\it left column\/}), and $\sigma_V = 10^{-1}$ ({\it right column\/}).
        \emph{Bottom row:} Density plots comparing the results obtained within each column. Each row of the 2D histograms has been normalized by the value of its integral.}
         \label{fig:test_I}
    \end{figure}
    
\subsection{Scaling of the problem}
The terms in Eq.~(\ref{chi_long}) mix quantities of a very different nature. The leftmost term is the usual definition of $\chi^2$ and if the noise is correctly estimated and the model can realistically represent the observed data, then that term should be of the order of one. The rightmost term depends on a scaling factor ($\alpha$) and the square of differences between values of the magnetic field, in Gauss units in our case. In order to make that term also close to one, we can add a constant that scales those terms so that the term itself is close to one. Our proposal is to use:
\begin{equation}
    \alpha' = \frac{\alpha}{4B_{\mathrm{norm}}^2},
\end{equation}
where $B_{\mathrm{norm}}$ is a typical norm for those magnetic field differences ($\approx 100$~G). If the problem is scaled this way, both terms should be close to unity and the optimal value of $\alpha$ can be usually found in the range $(0.1-10)$, depending on the target and how good the estimate of the norm is. For $B_\bot$, the norm must be squared twice as the WFA is linear with the square of the magnetic field. 

If the noise estimate is not included in the WFA least-squares problem, the two terms are unbalanced, usually requiring very small values of $\alpha$. In the following section, we do not include this scaling (so we assume $\sigma=1$ and $\alpha'=\alpha$) in order to be able to compare absolute numbers in our results, but with real data this scaling of the problem makes the search for the optimal $\alpha$ parameter trivial.
    
\section{Tests with an 3D rMHD simulation}\label{sec:mhd}

    In order to assess the performance of the spatially-regularized WFA, we have performed a simulated observation in the \ion{Ca}{ii}~$8542$~\r{A} line using a realistic 3D magnetohydrodynamics (MHD) simulation. We have taken one snapshot from a publicly available enhanced network simulation \citep{carlsson2016} performed with the \textsc{Bifrost} code \citep{gudiksen}. Snapshots from this simulation have been extensively used in previous studies and have been the workhorse of polarized line formation studies in recent years using the same spectral line \citep[e.g.,][]{stepan2016, quintero2017, jurcak2018, delacruz2019}.
    
    Since the height sensitivity of the 8542~\r{A} line is target-dependent and corrugated, the magnetic field inferred using the WFA might not correspond to a unique geometrical height in the simulation in all pixels, making the comparison with the input magnetic field vector non trivial. Since our goal is not to assess the effect of magnetic field gradients on the WFA, but the improvement delivered by our new method when compared to the traditional WFA, we have set the vertical stratification of the magnetic field in each pixel to a constant value, which corresponds to the vertical average of the magnetic field at the same pixel from $1000 \leq z \leq 1500$~km from the mean continuum formation layer. This situation is perhaps more idealized than what we would find in a normal observation where magnetic field gradients are expected as a function of height, however, many studies have overcome the effects derived from opacity jumps by restricting the wavelength range that is used to calculate the WFA \citep[e.g.,][]{jaime2013,delpinoaleman2016, kleint2017,sara2019}.
    
    
    We computed our synthetic observations using a modified version of the RH 1.5D code \citep{pereira2015,uitenbroek2001} that has been updated with a cubic DELO-Bezier solver of the polarized radiative transfer equation \citep{delacruz2013}.
   Figure~\ref{syn_stokes} presents monochromatic images in the four Stokes parameters at 80~m\AA\ from line center. The panels show a clearly chromospheric landscape with elongated magnetic features connecting two opposite-polarity patches.

    \paragraph{The effect of noise in Stokes $I$:} Before discussing our results, we would like to note that the WFA model relies on having a good estimate of $\partial I/\partial \lambda$, which is also derived from an observed quantity. If the observations are very noisy also in Stokes $I$, the estimate of $\partial I/\partial \lambda$ has a large uncertainty and that has an impact on the reconstructions, regardless of the noise that is present in $Q$, $U$ and $V$. We have performed a simple test in which we assume a large noise level of $\sigma=10^{-1}$ in $I$ only. We have reconstructed $B_\parallel$ twice using the traditional WFA algorithm, once with the noise added $I$ and a second time without adding noise. As Figure~\ref{fig:test_I} shows, the effect is clear and when the noise level in Stokes $I$ is high, the reconstructed magnetic fields have lower amplitudes than in the noiseless case (cf.~left column). The reason is probably that the correlation between $V$ and $\partial I/\partial \lambda$ starts to break and the numerator in Eq.~(\ref{eq:blos}) becomes smaller. We have repeated the same test but this time with the same amount of noise in Stokes~$V$ and we get identical results when we use the spatially-regularized algorithm (right column in Fig.~\ref{fig:test_I}).
    
    \paragraph{The line-of-sight component:} We have applied four different levels of noise $\sigma$ to the simulated data (relative to the continuum intensity), including $\sigma=0$. These values have been added to the data through a random Gaussian distribution. For each component of the magnetic field we have explored a range of $\alpha$ values that illustrate the effect of our method for each noise level. 

\begin{figure}
    \centering
    \includegraphics[width=\hsize]{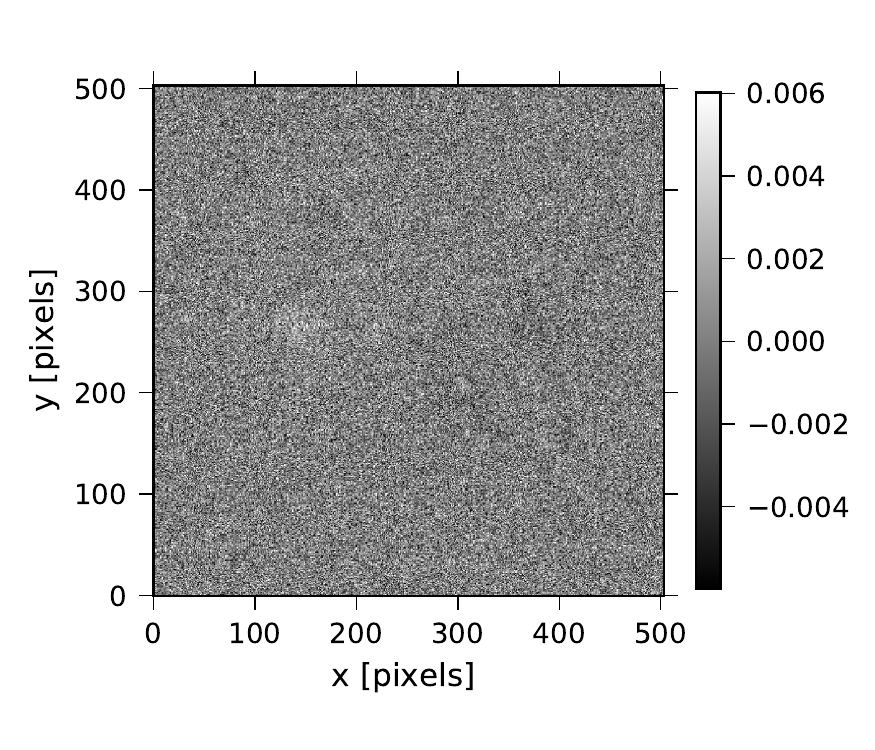}
        \caption{Monochromatic \ion{Ca}{ii}\,\,8542\,\AA\ Stokes $V$ image at $\Delta\lambda=118$~m\AA\ when the noise level is $\sigma = 10^{-1}I_c$. The signal is barely visible above the noise in some regions.
        }
         \label{stokesV_wave47}
    \end{figure}

    Depending on the choice of noise level $\sigma$, the signal might get drowned completely (in particular for $\sigma = 10^{-1}$) and we need to verify that our Stokes $V$ values are non zero.
    However, even for high noise levels we find non negligible signal in Stokes $V$ at wavelengths around line center, as
    Fig.~\ref{stokesV_wave47} shows for Stokes $V$ at $\Delta\lambda=118$~m\AA\ from line center. Here the signal is very weak, but with the spatially-regularized method we are nonetheless able to recover a non random map for $B_{\parallel}$.

    \begin{figure*}[!ht]
        \centering
        \includegraphics[width=0.9\textwidth]{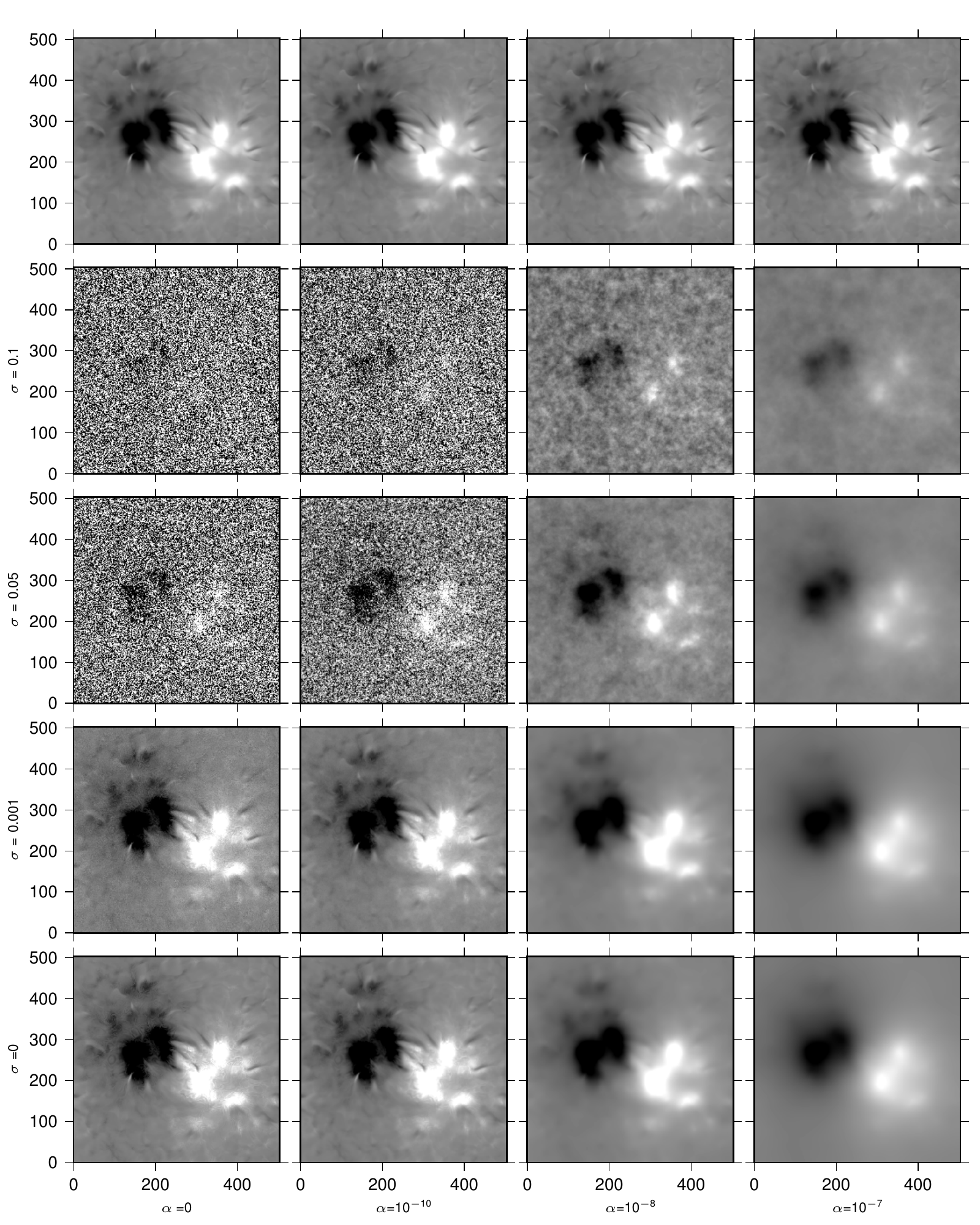}
        \includegraphics[width=0.9\textwidth]{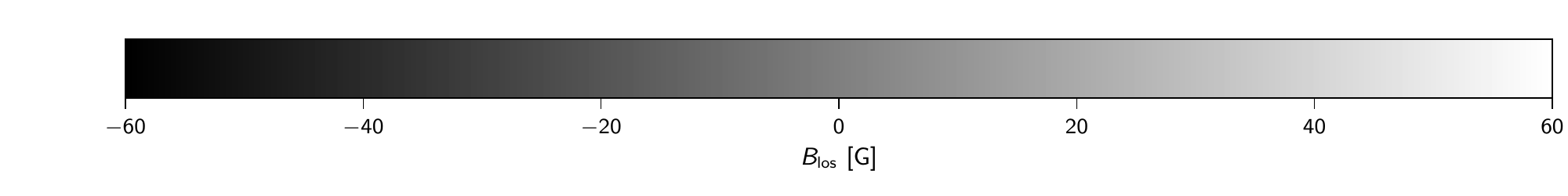}
        \caption{Inferred $B_{\parallel}$ from the spatially-regularized WFA of Bifrost synthetic data for different values of noise $\sigma$ and regularization parameter $\alpha$. 
        The top row shows the same $B_{\parallel}$ map from the MHD model for reference. 
        The remainder of the grid presents results for increased regularization parameter ({\it from left to right\/}) -- where the first column is the equivalent of the classical WFA -- and increased noise level $\sigma$ ({\it from top to bottom}).
        }
        \label{Blos_grid}
    \end{figure*}

    Figure~\ref{Blos_grid} presents the retrieval of the longitudinal component of \Bvec\ for the entire FOV obtained by imposing spatial constraints on the WFA. The grid shows all combinations of chosen regularization parameter values  $\alpha= [0, 10^{-10}, 10^{-8}, 10^{-7}]$ and noise levels $\sigma = [0, 10^{-3}, 5\times10^{-2}, 10^{-1}]$. For comparison, in the upper row of the grid we have plotted the constant magnetic field stratification that we used to synthesize the profiles.

    \begin{figure*}[ht]
        \centering
        \includegraphics[width=15.2cm,height=15.2cm]{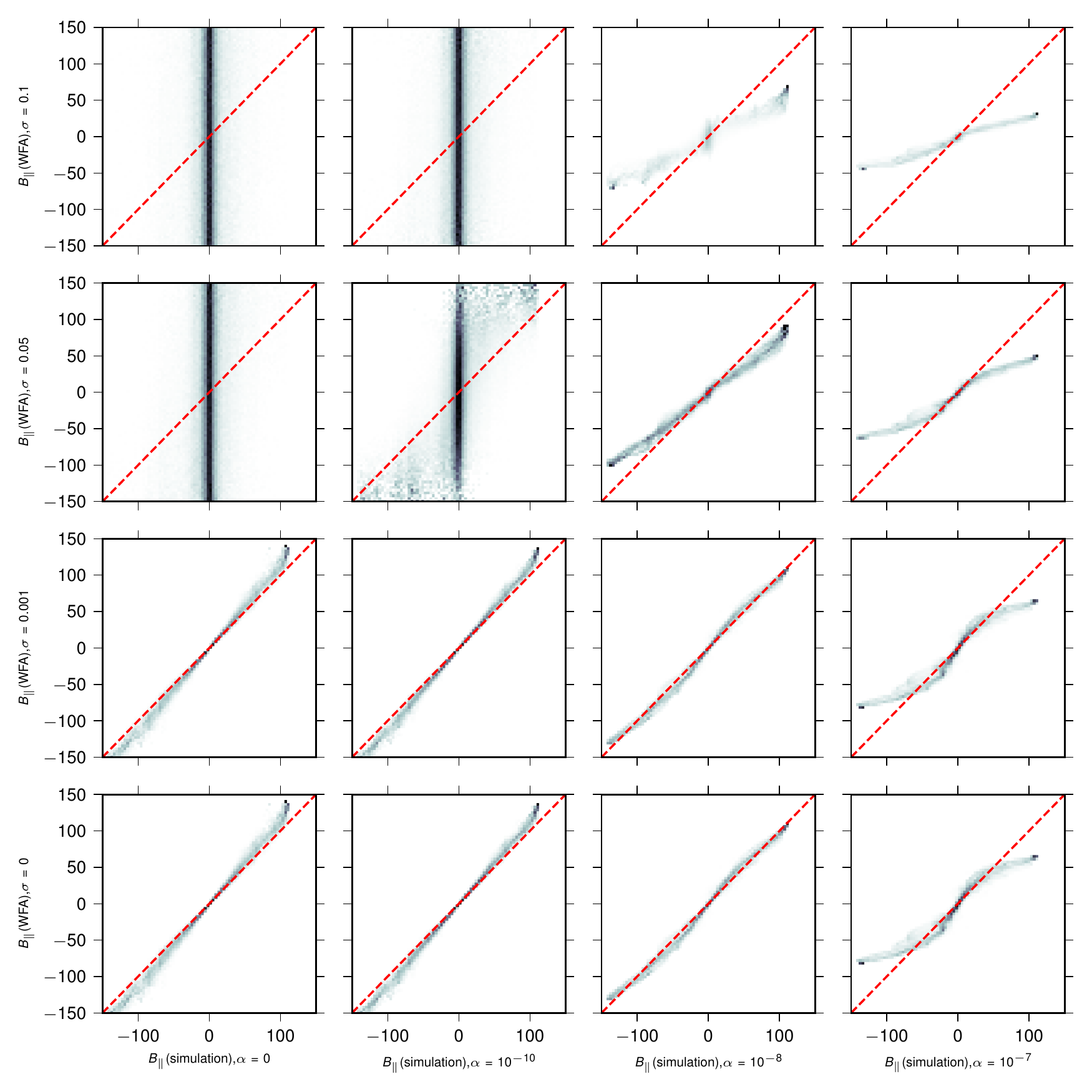}
        \caption{Density maps comparing the values for $B_{\parallel}$ from the MHD simulation ($x$-axis) with those from the spatially-\textbf{regularized} WFA ($y$-axis). The panels are arranged in the same order as the ones in the grid of Fig.~\ref{Blos_grid}. Each column in each plot has been normalized by the sum of the values in that column.}
        \label{Blos_dens}
    \end{figure*}
    
    When the noise levels are low (bottom rows), the amplitude and shape of the Stokes profiles is well-determined and very little or no regularization is required (lower-left part of the grid). As we increase the noise level, the solution becomes more affected by noise, but using spatial constraints allows diminishing the impact of noise in the maps, while keeping the amplitudes of the reconstructed magnetic field patches. In the extreme case of $\sigma = 10^{-1}$ (second row) there is a hint of the original structure of the magnetic field in the non regularized WFA but the amplitude of the noise is as large as those signals. Increasing the regularization parameter reduces the noise level and a coherent structure can be recovered. We note however that the reconstruction is not perfect and the regions with very weak magnetic field are lost as they did not leave an imprint above the noise in the Stokes profiles. This imprint does not need to be large at all, and even a very small hint of signal above the noise level seems to be enough (see Fig.~\ref{stokesV_wave47}). Overall, the optimal $\alpha$ parameter for each noise level seems to be located around the counter-diagonal of the grid.
    
    Figure~\ref{Blos_dens} presents 2D histograms of the reconstructed magnetic field amplitudes between the MHD simulation values and each case of the grid, allowing a quantitative comparison.
    Overall, these scatter plots show that when the noise is low and the $\alpha$ parameter is not overestimated (lower left six panels), both the traditional and the spatially-regularized reconstructions follow the one-to-one line. When the $\alpha$ parameter is overestimated, the spatially-constrained method does not allow for a sufficiently rapid spatial variation in the FOV and the strongest field amplitudes are suppressed.
    
    When the noise level is large and comparable to the signal amplitudes, the traditional method performs poorly and only a noisy image is reconstructed (top left four panels). In this case the spatially-regularized method clearly outperforms the traditional method by using information from surrounding pixels. The scatter plots corresponding to $\sigma=10^{-1}$
    show that even for an $\alpha$ value that was not overestimated for lower noise levels (third column), the reconstructed magnetic field has lower amplitudes than the original.
    We have investigated this effect and found that it partly originates from the fact that Stokes~$I$ is very noisy and therefore its derivative is also poorly determined observationally (see Fig.~\ref{fig:test_I} and discussion thereof). We have conducted a test where we did not add the noise to Stokes~$I$ and the reconstructed amplitudes were much closer to the one-to-one reconstruction. Having a noise level of $\sigma=10^{-1}$ is very unlikely in a real observation.

    \begin{figure*}[ht!]
    \centering
        \includegraphics[width=0.92\textwidth]{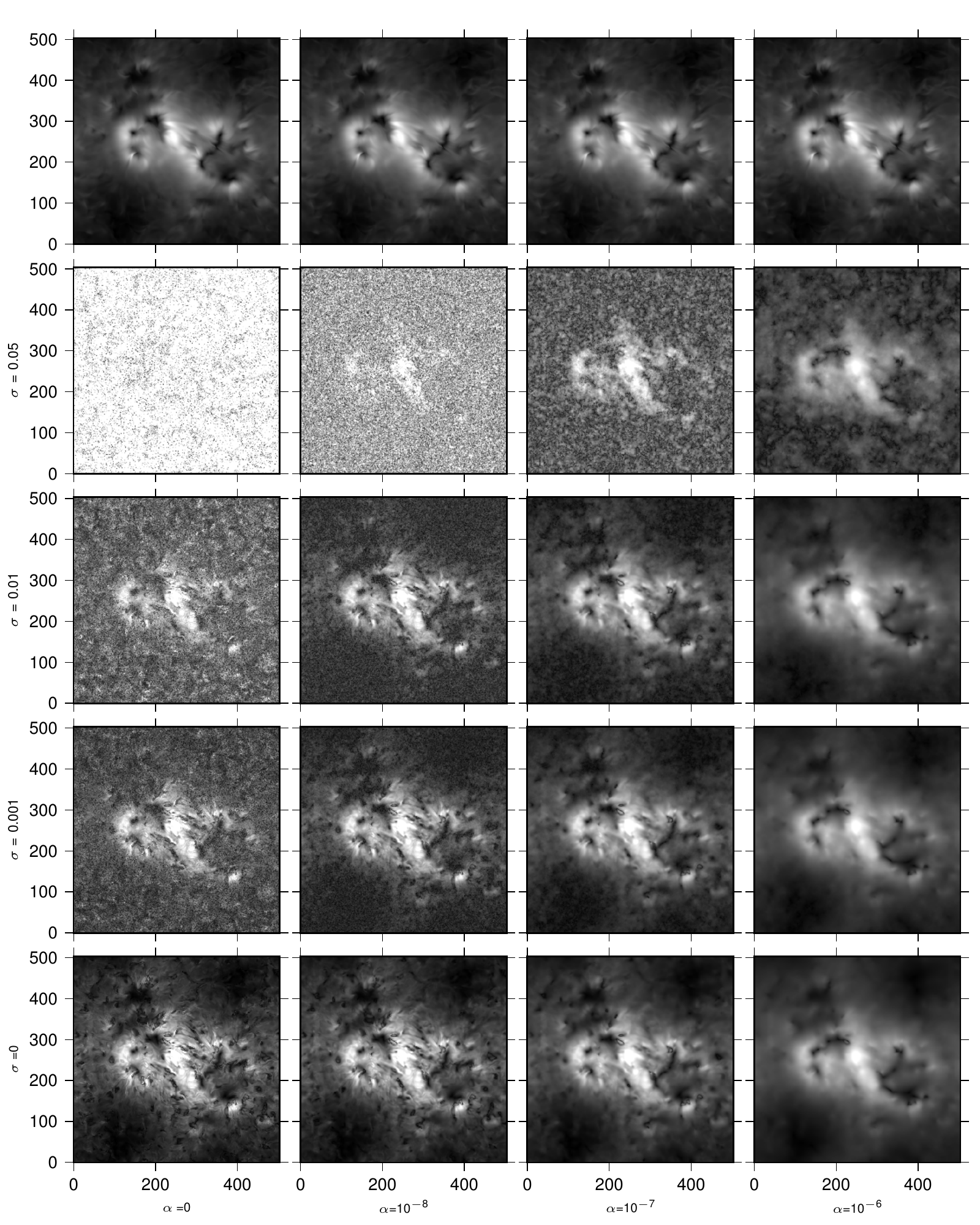}
        \includegraphics[width=0.92\textwidth]{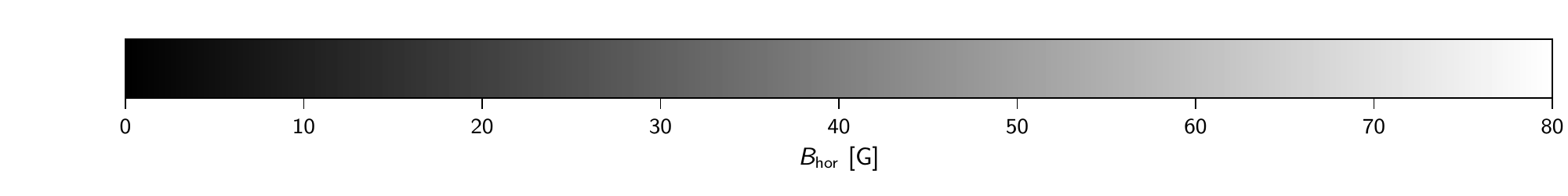}
        \caption{Inferred $B_{\bot}$ for different values of noise $\sigma$ and regularization parameter $\alpha$. 
        Format as for Fig.~\ref{Blos_grid}.
        }
    \label{Bhor_grid}
    \end{figure*}
    
    \begin{figure*}[ht!]
        \centering
        \includegraphics[width=15.2cm,height=15.2cm]{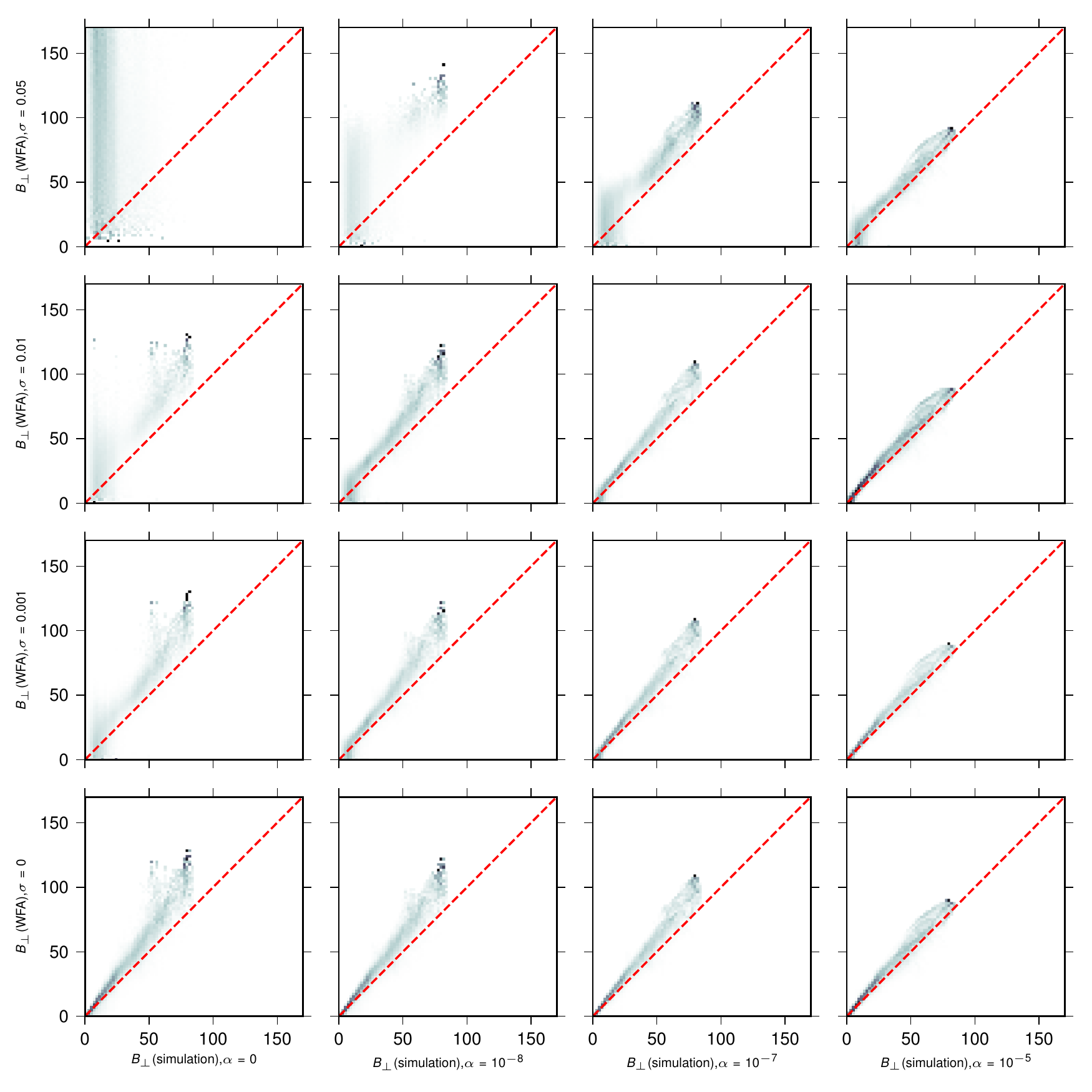}
        \caption{
        Density maps comparing the values for $B_{\bot}$ from the MHD simulation ($x$-axis) with those from the spatially-\textbf{regularized} WFA ($y$-axis).
        Format as for Fig.~\ref{Blos_dens}.
        }
        \label{Bhor_dens}
    \end{figure*}
    
    \paragraph{The transverse component of the magnetic field:} 
    Since the amplitudes of the signals in $Q$ and $U$ are smaller than in $V$, the range of $\alpha$ parameter values has been adjusted accordingly when computing the transverse component of the magnetic field. Specifically, we have chosen $\alpha=[0, 10^{-8}, 10^{-7}, 10^{-6}]$. The noise levels in this case are $\sigma = [0, 10^{-3}, 10^{-2}, 5 \times 10^{-2}]$. Using the same regularization values as for $B_{\parallel}$, it would not have been possible to obtain $B_{\bot}$ for the largest noise levels. 
    
    Figure~\ref{Bhor_grid} presents the results obtained for $B_{\bot}$. The first row corresponds to the horizontal component of magnetic field from the MHD simulation. The effect of adding spatial constraints in this case yields similar results to the reconstructions of $B_\parallel$, but with some notable differences. When the noise level is low or absent (lower-left corner of the grid) we would expect the WFA to yield an almost perfect reconstruction, however this is not the case. There are several reasons why the result is actually showing more spatial structure (artifacts) than the top row:
    \begin{itemize}
        \item The velocity field present in the simulation is affected by gradients that can make the WFA not valid for the transverse component of $\boldsymbol{B}$, in particular when those gradients are strong.
        \item Eq.~(\ref{wingsq}) assumes that we know the wavelength offset relative to the local frame of rest of the atmosphere where the signal is originating. So even if there are no velocity gradients, a constant velocity causes artifacts. 
        \item The WFA for Stokes $Q$ and $U$ is a second-order term and its applicability is limited to very weak fields, which might not always be the case in these synthetic data.
    \end{itemize}
    Although all these factors might be acting simultaneously, we can try applying the WFA, carefully assessing the quality of the results. Since in a real observation we have no prior knowledge of the velocity stratification, we have not attempted correcting the wavelength grid in each pixel for the chromospheric line-of-sight mean velocity. As a result, even for low noise levels, having a larger regularization parameter helps removing those artifacts. When the noise level is very large, the traditional WFA reconstruction yields only noise (second row, leftmost panel). As the regularization parameter is increased, the algorithm is capable of lowering the noise while recovering more structure from the data. 

    Figure~\ref{Bhor_dens} presents 2D histograms of original versus reconstructed $B_{\bot}$, in similar format as Fig.~\ref{Blos_dens}.
    We find that the unconstrained results overestimate the magnetic field amplitude due to the presence of small-scale artifacts. Therefore, for all cases using a larger $\alpha$ than what a low noise level would otherwise require seems to produce the correct amplitudes (see the rightmost column in Fig.~\ref{Bhor_dens}). While we unfortunately have no way to know the real solution {\it a priori\/} from observations, a good approach would seem to be to choose the smallest value of $\alpha$ that yields a smooth reconstruction at the smallest scales.

    \begin{figure*}[ht!]
        \centering
        \includegraphics[width=0.92\textwidth]{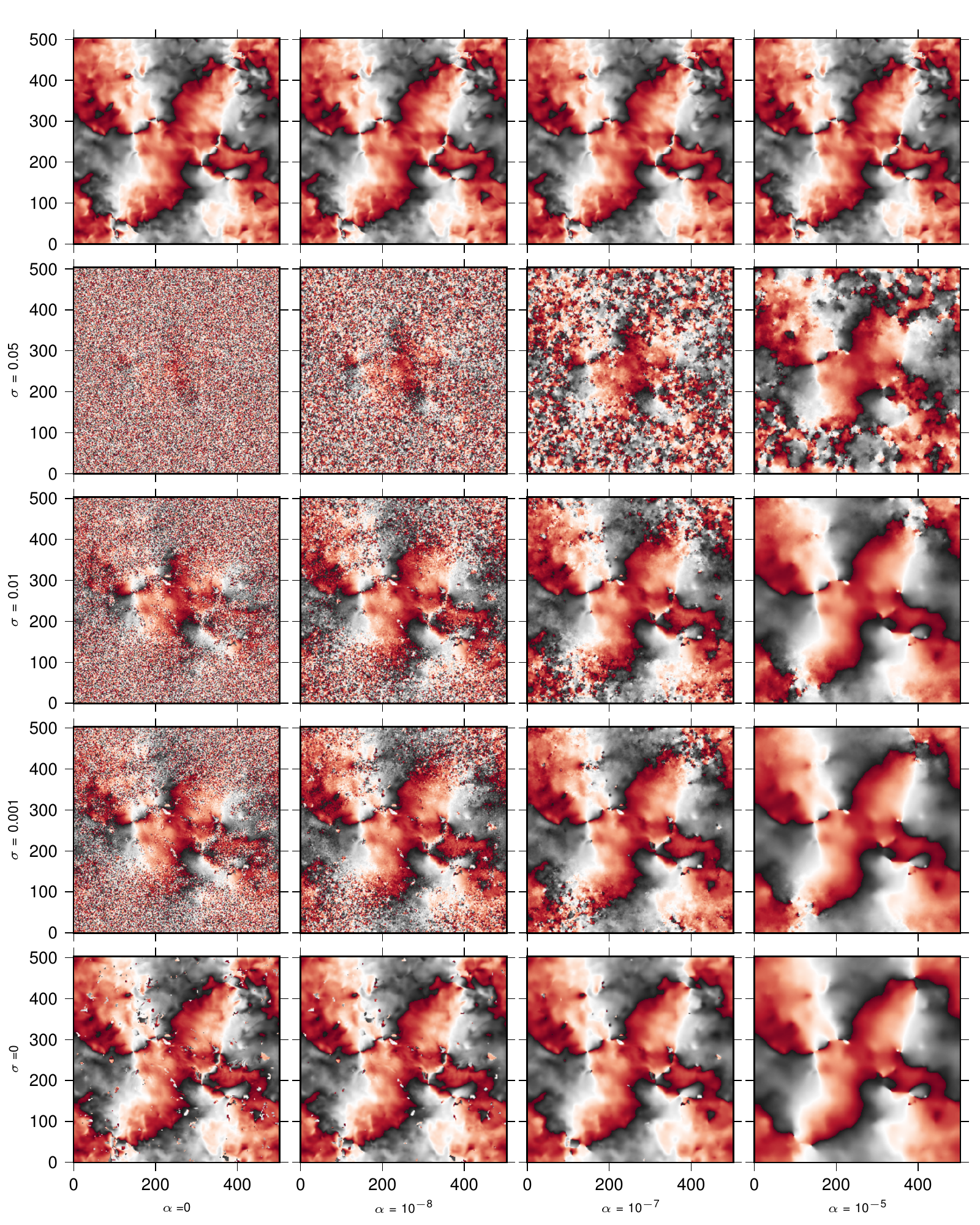}
        \includegraphics[width=0.92\textwidth]{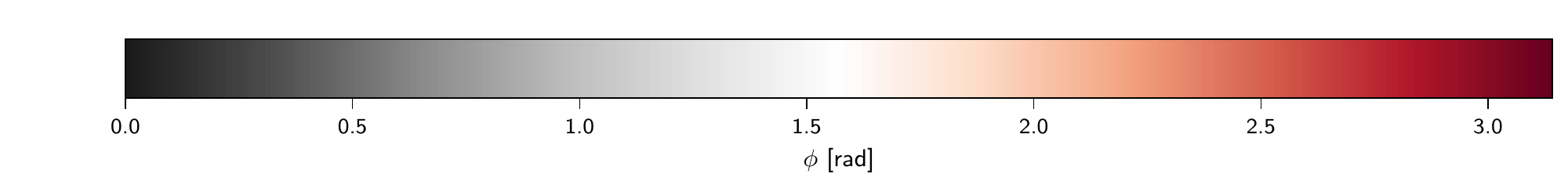}
        \caption{Inferred $\phi$ for different values of noise $\sigma$ and regularization $\alpha$. 
        Format as for Figs.~\ref{Blos_grid} and \ref{Bhor_grid}.
        }
        \label{azim_grid}
    \end{figure*}
    
    \begin{figure*}[ht!]
        \centering
        \includegraphics[width=15.2cm,height=15.2cm]{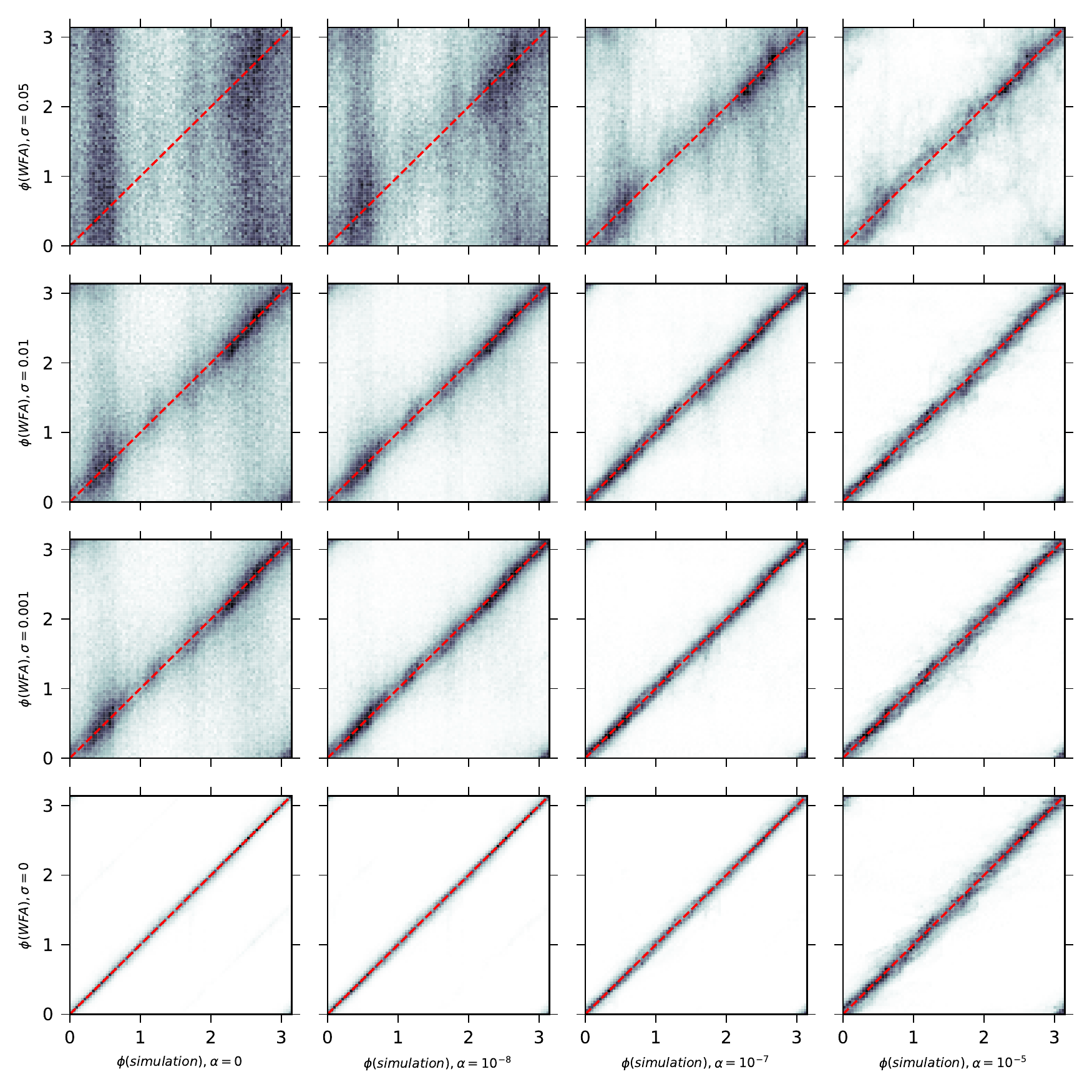}
        \caption{
        Density maps comparing the values for $\phi$ from the MHD simulation ($x$-axis) with those from the spatially-\textbf{regularized} WFA ($y$-axis).
        Format as for Figs.~\ref{Blos_dens} and \ref{Bhor_dens}.
        }
        \label{azim_dens}
    \end{figure*}
    
    \paragraph{The magnetic field azimuth:} The same procedure as been repeated for $\phi$ using the formulae obtained in section \ref{azimuth}. The regularization and noise values are the same as in the case of the horizontal component of \Bvec, that is to say, $\alpha=[0, 10^{-8}, 10^{-7}, 10^{-6}]$ and $\sigma = [0, 10^{-3}, 10^{-2}, 5 \cdot 10^{-2}]$. Figure~\ref{azim_grid} presents the reconstructed grid for the azimuth for the entire FOV. For comparison we show the azimuth from the MHD simulation in the upper row. Again we find that applying WFA without any spatial constraints does not give reliable results if the level of noise is non zero. However, if we increase the weight that the nearby pixels have, we see that the map for the azimuth becomes much more similar to the predicted ones. 
    
    In the case of the azimuth it is clear that the Stokes $Q$ and $U$ signal outside the central part of the FOV is very weak, but given the very smooth nature of the azimuth we can choose a relatively large $\alpha$ value that tightly constrains the largest scales. Even for the largest noise level that we have considered, the reconstruction of the azimuth is very consistent spatially with the MHD map, although the errors in the outer part of the FOV are noticeably larger.
    
    With real observations we can optimize the regularization parameter $\alpha$ by exploring a range of $\alpha$ values (which then is not a grid since the noise level is fixed) and again selecting the smallest $\alpha$ for which the reconstruction is smooth also at small scales.
    
    In Appendix~\ref{ap:1} we discuss other benefits of imposing regularization that we found when analyzing real observations.

\section{Magnetic fields in active region plage}
\subsection{Data acquisition and reduction}
    Our observations were acquired at the Swedish 1-m Solar Telescope \citep[SST;][]{scharmer2003} on 2018-06-19 at 07:30:46--07:43:43~UT, targeting a plage region at coordinates ($X$,$Y$) = (229\arcsec,61\arcsec), corresponding to viewing angle $\mu=0.97$. The CRisp Imaging Spectro-Polarimeter \citep[CRISP][]{scharmer2008} obtained full-Stokes observations in the \ion{Ca}{ii}~8542~\AA, \ion{Na}{i}~5896~\r{A} and \ion{Mg}{i}~5173~\AA\ lines.
    
    \ion{Ca}{ii}~8542~\AA\ was critically sampled at 21 equidistant positions around the line center with a step size of $\Delta\lambda = 55$\,m\AA. Two extra points in the outer wings are located at $\pm 0.88$\r{A} from the line center. The \ion{Mg}{i} line was sampled at 13 equidistant positions $\pm 0.04$ \r{A} around line center and the two wing points located at $\pm 0.24$\r{A} from the center of the line.  Finally, \ion{Na}{i}~5896 \r{A} line was sampled at 13 positions around line center with a step size of $\Delta \lambda = 60$~m\AA, with an extra wing point located at $-0.6$~\AA\ from line center. The CRISP data were sampled with a pixel scale of 0\farcs{059} and the total cadence is 37\,s. 
    
    The data were reduced with the CRISPRED/SSTRED data pipeline \citep{jaime2015,lofdahl,2011A&A...534A..45S}. Compensation of atmospheric seeing effects was performed using the Multi-Object Multi-Frame Blind Deconvolution method \citep[MOMFBD;][]{vannoort,lofdahl2002}. All three datasets were co aligned and corrected for residual rubber-sheet distortions using local cross-correlation of small patches in the image where the \ion{Ca}{ii} dataset was used as a reference.
    
    \begin{figure*}[!ht]
        \centering
        \includegraphics[width=0.88\textwidth]{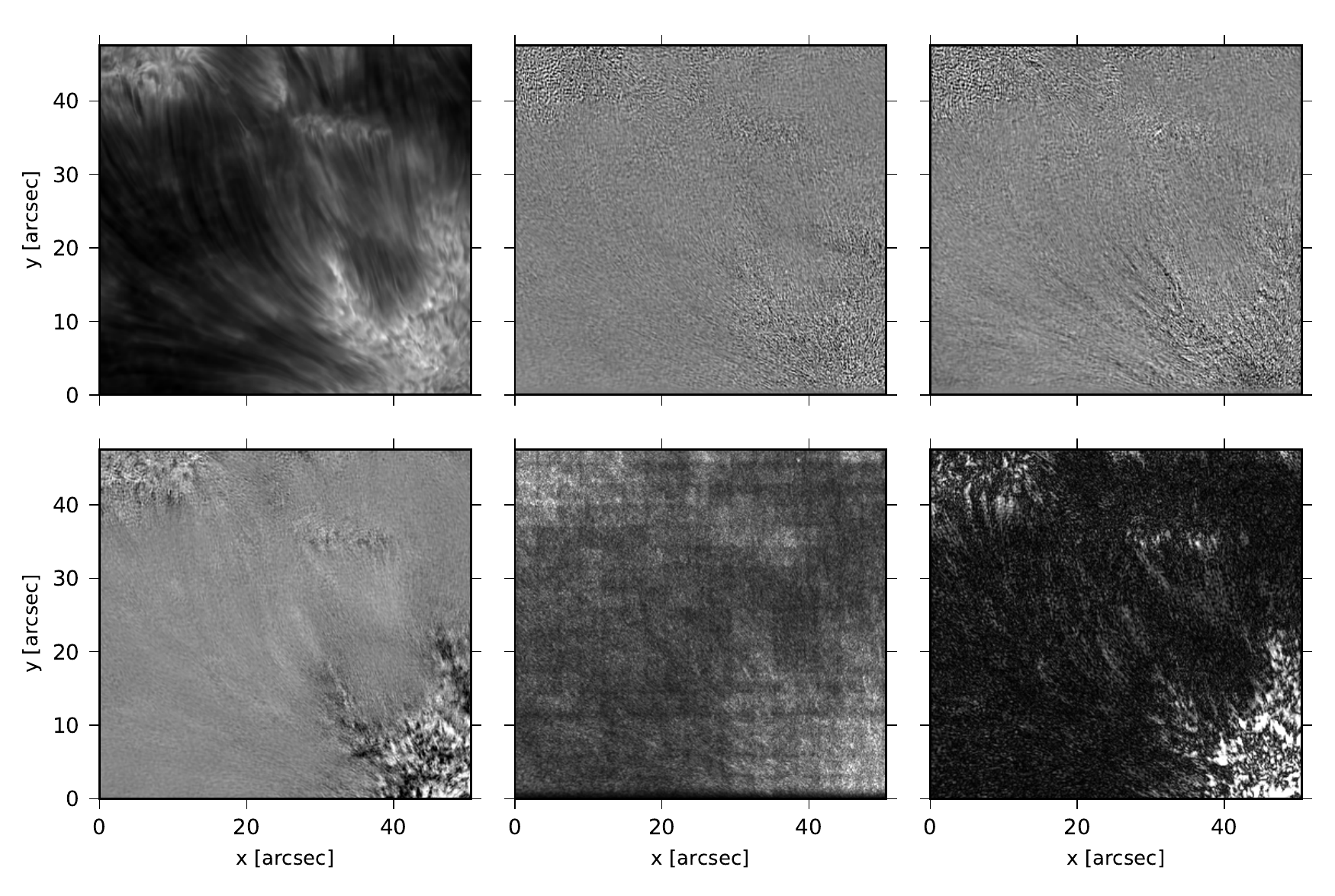}
        \caption{
        Overview of the \ion{Ca}{ii}~8542~\AA data at $\Delta \lambda = $0.036 \r{A} from line center.
       \emph{Top:} Stokes $I$,  $Q$ and $U$. \emph{Bottom:} Stokes $V$, wavelength-averaged linear polarization map based on Stokes $Q$ and $U$, and  wavelength-averaged circular polarization map based on Stokes $V$. }
        \label{stokes_data_ca}
    \end{figure*}

\subsection{Overview of the observations} 
\begin{figure*}[!ht]
        \centering
        \includegraphics[width=0.88\textwidth]{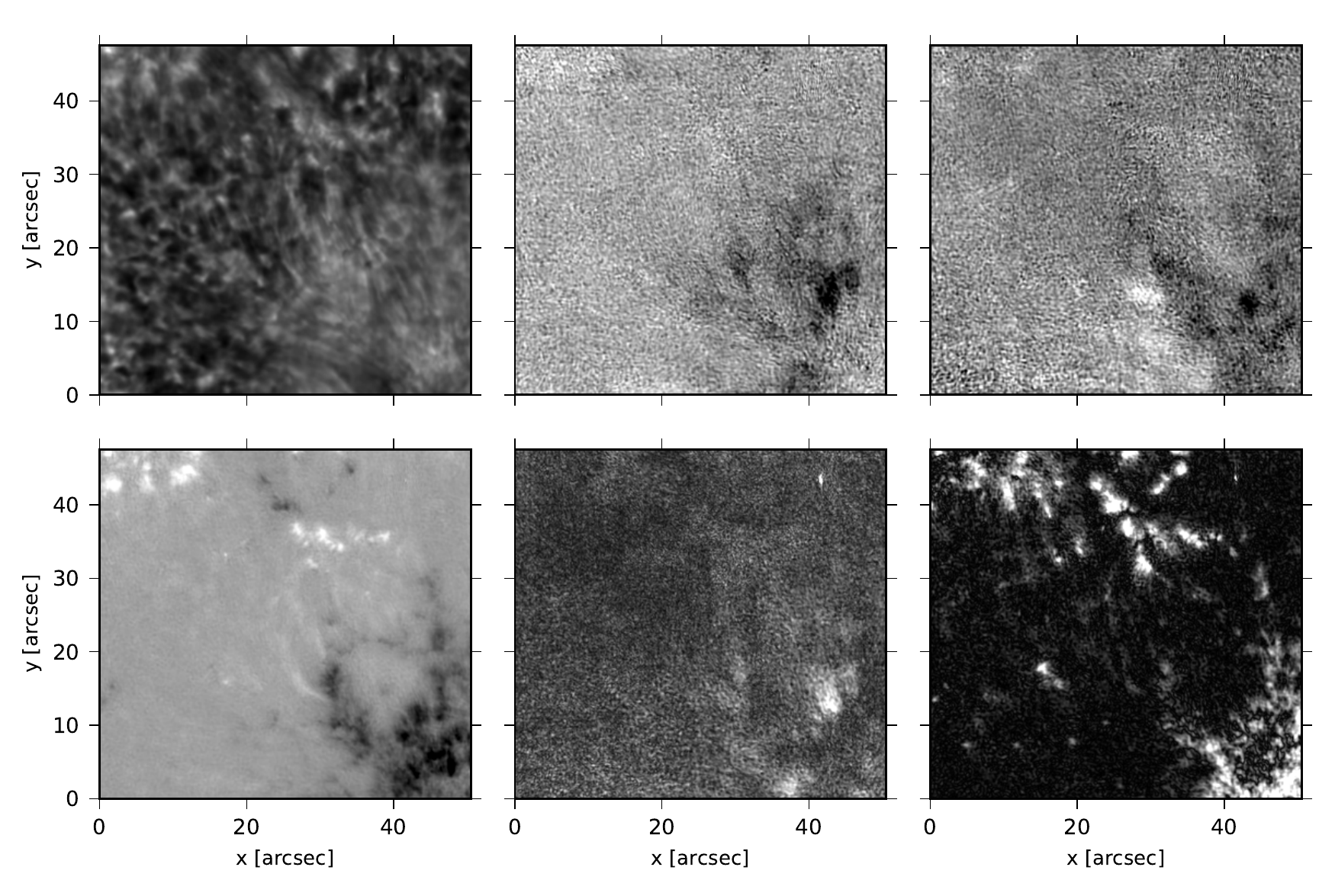}
        \caption{
        Overview of the \ion{Mg}{i}~5173~\r{A} data at $\Delta \lambda = $0.040 \r{A} from line center.
        Format as for Fig.~\ref{stokes_data_ca}.
        }
        \label{stokes_data_mg}
    \end{figure*}

\begin{figure*}
        \includegraphics[width=0.88\textwidth]{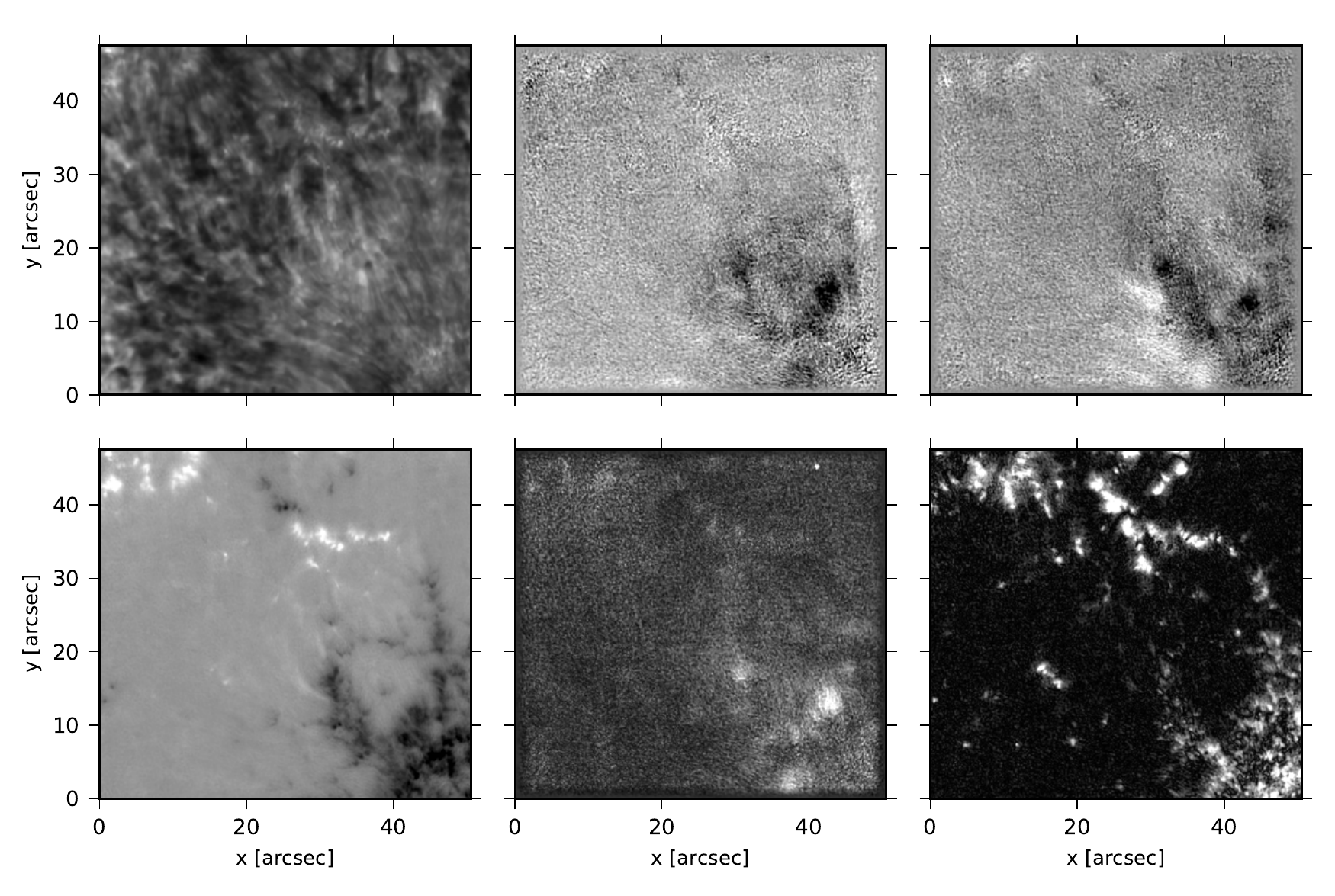}
        \caption{
        Overview of the \ion{Na}{i}~5896 \r{A} data at $\Delta \lambda = $0.040 \r{A} from line center.
        Format as for Fig.~\ref{stokes_data_ca} and \ref{stokes_data_mg}.
        }
        \label{stokes_data_na}
    \end{figure*}

    We analyze a single scan taken at 07:33:14 UT, which had the best seeing conditions of the time-series. Figures \ref{stokes_data_ca}, \ref{stokes_data_mg} and \ref{stokes_data_na} show an overview of the observations in \ion{Ca}{ii}~8542~\AA, \ion{Mg}{i}~5173~\AA\ and \ion{Na}{i}~5896~\r{A}, respectively. The maps are not exactly at line center but at $\Delta \lambda =$ 55 m\r{A} in the \ion{Ca}{ii} line, $\Delta \lambda =$ 40 m\r{A} in the \ion{Mg}{i} line and $\Delta \lambda =$ 60 m\r{A} in the \ion{Na}{i} line, where more signal can be expected in Stokes~$Q$, $U$ and $V$ than at line center. 
    
    In the Stokes~$V$ maps (bottom left) two regions with opposite polarities are visible in the upper-left and lower-right corners of the image. These regions are connected by elongated, fibrilar features evident in the \ion{Ca}{ii} Stokes~$I$ panel (Fig.~\ref{stokes_data_ca}, top left). 
    The signal, especially in Stokes $Q$ and $U$ (top middle and right), is stronger and above the noise level in the \ion{Mg}{i}~5173~\AA\ and \ion{Na}{i}~5896~\r{A} lines, while for \ion{Ca}{ii}~8542~\AA\ $Q$ and $U$ are weak and at the detection and calibration limit. Stokes $V$ exhibits signal above the noise level for all three lines. 
    One way of improving the signal-to-noise would be by stacking several scans through time, however, in our case we did not consider this to be a good option because of the cadence of the observations (over half a minute) and the presence of waves in the solar atmosphere. 
    
    The bottom middle and right panels in Figs.~\ref{stokes_data_ca}--\ref{stokes_data_na} show wavelength-averaged linear and circular polarization maps, based respectively on Stokes $Q$\&$U$ and $V$. They were computed using only the central wavelength points -- five for \ion{Ca}{ii} and three for \ion{Mg}{i} and \ion{Na}{i} -- and saturated to increase the visibility of the structures present in the images. Finally, we note that the grid in the linear polarization map of \ion{Ca}{ii} (Fig.~\ref{stokes_data_ca}) is an artifact of the mosaicing technique used in the data reduction process. 

    \subsection{Sensitivity of the observed spectral lines}\label{sec:sens}  

    \begin{table*}
    \centering
     \begin{tabular}{c |c |c | c | c| c | c | c | c} 
     \hline\hline
     \centering
        Label &    Atom & Line [\AA] & $\bar{g}_{\mathrm{eff}}$ & G & $\lambda_i$ [m\AA] & $<z>$ [Mm] & $<B_\parallel>$ [G] & $<{B}_\parallel>$ [G] (2D)\\
            \hline
%
    a) & \ion{Na}{i} & 5895.824 & 1.33 & 1.33 & $[-360,-300,-240,240,300,360]$ & $157 \pm 18$ & $398 \pm 206$ & $719 \pm 88$\\
    b) & \ion{Na}{i} & 5895.824 & 1.33 & 1.33 & $[-120,60,60,120]$ & $474 \pm 40$ & $546 \pm 161$ & $692 \pm 94$\\
    c) & \ion{Mg}{i}  & 5172.684 & 1.75 & 2.87 & $[-40,0,40]$ & $760 \pm 45$ & $502 \pm 118$ & $558 \pm 103$\\         
    d) & \ion{Ca}{ii} & 8542.091 & 1.10 & 1.21 & $[-110,-55,0,55,110]$ & $1168\pm 122$ & $417 \pm 69$ & $444 \pm 83$\\
     \hline
    \end{tabular}
    \caption{Properties of the spectral windows considered in this study. $<z>$ and $<B_\parallel>$ are spatial averages that have been computed from the slit marked in Fig.~\ref{fig:Blos_slice}. We have also indicated the standard deviation of the spatial variations. $<{B}_\parallel>$ corresponds to the mean of $B_\parallel$ within each window computed over the bottom right plage patch, excluding field free gaps (yellow contours). The effective Land\'e factor $\bar{g}_{\mathrm{eff}}$ scales the amplitude of $V$ and $G$ scales the amplitude of $Q$ and $U$. $\lambda_i$ indicates the line positions relative to line center that have been used to apply the WFA.}             
    \label{table:1}
    \end{table*}


Our observations include three strong spectral lines that sample a vast range of heights of the solar photosphere and chromosphere. Due to opacity variations, the outer wings of these lines are sensitive to the photosphere, whereas the inner cores are sensitive to the chromosphere.
    
The \ion{Mg}{i}~5173~\AA\ and \ion{Na}{i}~5896~\AA\ lines share some similarities. Both are strong lines that sample a fuzzy boundary between the upper photosphere and the lower chromosphere. They both correspond to minority species and therefore they are strongly affected by scattering close to line center. In fact the Stokes~$I$ images in Figs.~\ref{stokes_data_mg} and \ref{stokes_data_na} are strikingly similar in appearance. The \ion{Mg}{i} line is more sensitive than the \ion{Na}{i} to Zeeman-induced polarization (see Land\'e factors in Table~\ref{table:1}). For more information about these lines we refer the reader to the publications by \citet{2011A&A...531A..17R} and \citet{2018MNRAS.481.5675Q}, and references therein.
    
The core of the \ion{Ca}{ii}~8542~\AA\ line samples the lower/middle chromosphere, and compared to the other two lines that we have discussed, it is more strongly coupled to the local physical conditions, which makes it also a better temperature diagnostic. The Stokes~$I$ panel in Fig.~\ref{stokes_data_ca} is dominated by chromospheric elongated feature that connect bright magnetically active patches. This line is the least sensitive to Zeeman-induced polarization of our sample. For more information about the 8542 line we refer the reader to previous studies \citep[e.g.,][and references therein]{2000ApJ...544.1141S,pietarila2007,2008A&A...480..515C,2009ApJ...694L.128L,2010ApJ...722.1416M,2012A&A...543A..34D,stepan2016}. 

In order to extract the stratification of the magnetic field, we can use the opacity variations within these lines to our advantage.
We have defined four different spectral windows that sample four different height regions distributed throughout the photosphere and lower chromosphere. 

\begin{figure*}[!ht]
    \centering
     \includegraphics[width=0.82\textwidth]{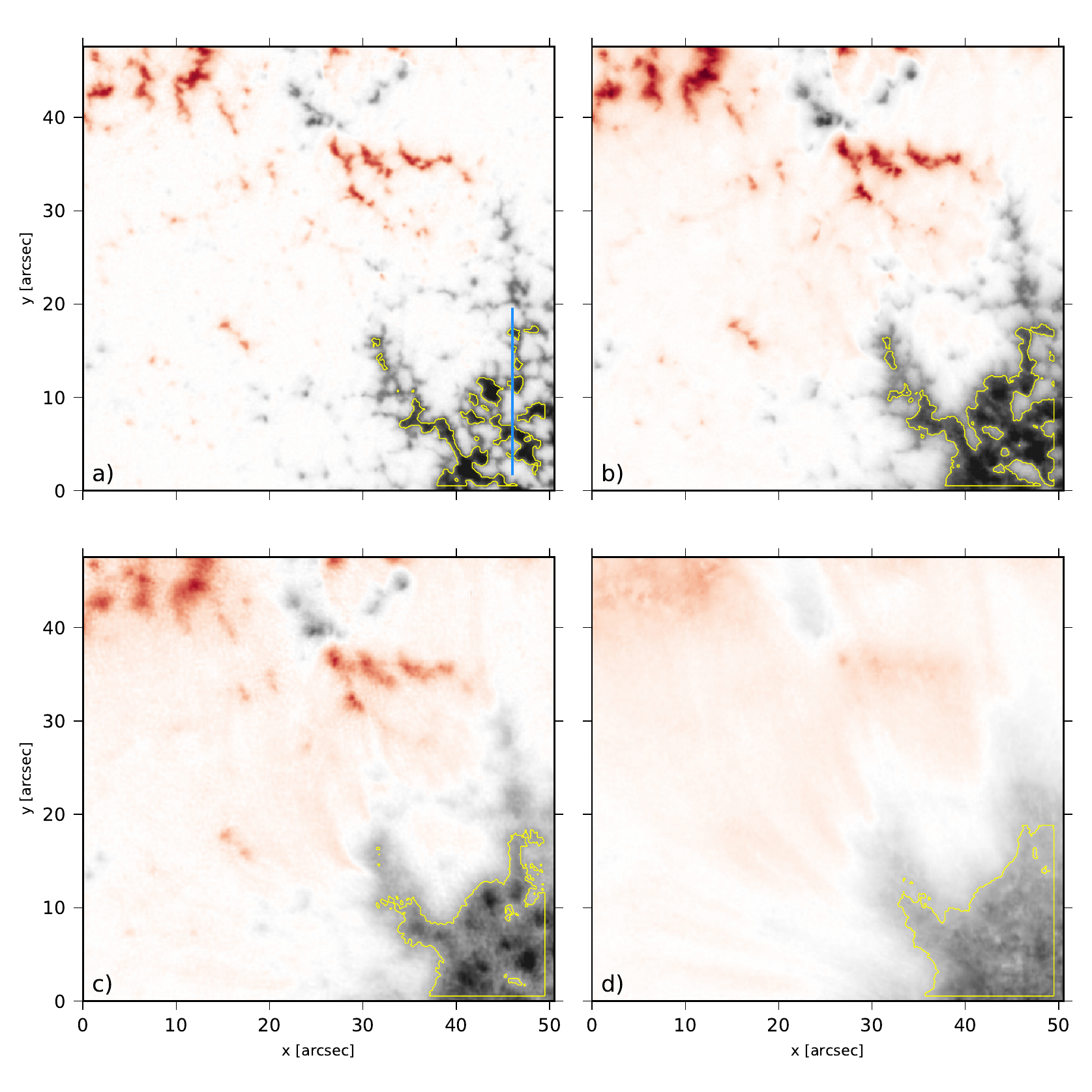}
     \includegraphics[width=0.82\textwidth]{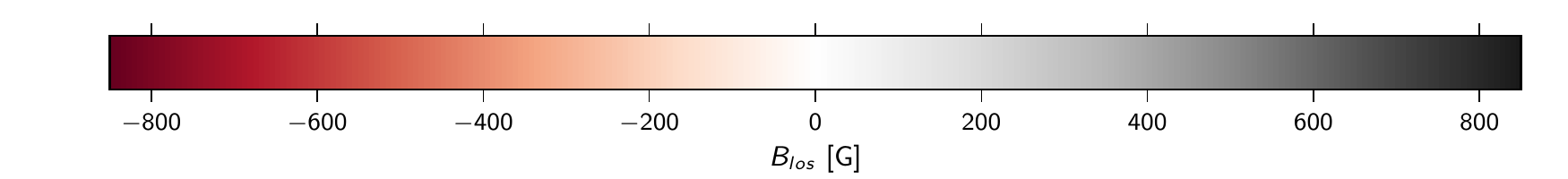}
    \caption{Inferred maps of $B_\parallel$ from the four spectral windows summarized in Table~\ref{table:1}. The panels are ordered according to their average formation height, as estimated using response functions. \emph{Panel (a):} Photospheric estimate derived from the far wings of the \ion{Na}{i}~5896~\AA\ line. \emph{Panel (b):} Upper photospheric estimate derived from the inner wing of the \ion{Na}{i}~5896~\AA\ line. \emph{Panel (c):} Lower chromospheric estimate derived from the core of the \textbf{\ion{Mg}{i}~5173~\AA\  line.} \emph{Panel (d):} Chromospheric estimate derived from the core of the \ion{Ca}{ii}~8542~\AA\ line. The yellow contours indicate the area used to compute the spatial averages that are listed in the rightmost column in Table~\ref{table:1}.}
    \label{Blos_all}
\end{figure*}

In order to quantify to what heights these four windows correspond, we have extracted the spectra from a slit indicated in Fig.~\ref{Blos_all} and performed a NLTE inversion of the three lines included in our observations using the STiC code \citep{delacruz2019}. We have used the output model from that inversion to calculate response functions \citep[RF; see review by][]{2016LRSP...13....4D}, which allow for an analysis of the sensitivity of each line to perturbations applied at each depth-point of the model. By utilizing the RFs from the model and the z-scale that is calculated during the inversion (in hydrostatic equilibrium) we can easily assign an approximate geometrical formation height to each of our spectral windows. 

The formation height of each spectral window has been estimated using the center of gravity of the RFs within the wavelength window that was used to calculate each panel. To derive the zero point of the geometrical scale we have calculated the mean gas pressure along the slit at the height where $\tau_{500} = 1$, including the effect of magnetic pressure, and we have shifted the $z$-scale of each column so that there is horizontal gas pressure equilibrium at the continuum formation layer. This approach is only an approximation but it is the best we can currently do without including the Lorentz force and magnetic pressure in the inversion process, which is not a trivial task \citep{2010ApJ...720.1417P,2019A&A...629A..24P,2019A&A...632A.111B}.

Some properties from the four spectral regions are summarized in Table \ref{table:1}, where we have indicated the wavelength range along with the spectral line properties for each window. The mean formation height of each region, from the deepest one to the highest one is: $(157, 474, 760, 1168)$~km from the surface, although we have knowledge of the individual values for each column within the slit. 

The \ion{Na}{i} and \ion{Mg}{i} lines included in this study reach similar heights at line center. However, our coarser sampling of the \ion{Na}{i} line makes it less optimal than the \ion{Mg}{i} line to use the line center as a chromospheric window due to strong opacity effects as a function of wavelength. Therefore the \ion{Na}{i} was used to estimate the magnetic field in the photosphere at two different heights. The better sampling achieved in the \ion{Mg}{i} line makes it better suited for an estimate using spectral positions close to line center while avoiding strong opacity effects, providing an estimate at the base of the chromosphere. The window that includes the \ion{Ca}{ii} line samples the highest height range.
 
\section{Results}

    
    

\subsection{Reconstruction of $B_\parallel$}\label{sec:result_blos}
We have applied the spatially-regularized WFA to the data in the four spectral regions as defined in Table~\ref{table:1} and derived two-dimensional maps of $B_\parallel$. The $\alpha$ parameter was calibrated using a grid similar to those displayed in Section~\ref{sec:mhd}. The optimal value was selected in such a way that noise was reduced compared to the unregularized WFA while not degrading the spatial structures that are present in the image.

In the discussion of our results we focus on the positive polarity patch in the lower-right corner of the FOV. In the deepest photospheric panel (Fig.~\ref{Blos_all}a) the magnetic field is spatially concentrated, interspersed with apparently field-free gaps (i.e., white in the maps). The line-of-sight field strength reaches kG values in most of these concentrations with a mean value (excluding the field-free gaps) of 719\,G (within the yellow contours). In panel (b) the mean formation height is estimated to be 200\,km higher up and the magnetic field has expanded and filled up a large fraction of the field-free gaps. The average strength of the field is now 692\,G. In panel (c) the magnetic canopy has almost completely formed, filling most of the FOV. The mean magnetic field strength has decreased to 558\,G. 
In this window we also estimate the horizontal component, which we discuss in more detail in Section~\ref{sec:result_bhor} below. In the final window the canopy has completely formed filling all the plage patch with an average canopy value of 444\,G. Most of the footpoint magnetic elements that were still visible in panel (c) are now hidden below the canopy. The clearest example of this effect can be found in the negative polarity patches located around $(x,y) = (10\arcsec,43\arcsec)$. The overall topology is smooth over the entire FOV.
 
  \begin{figure*}[!ht]
    \centering
    \includegraphics[width=0.9\textwidth]{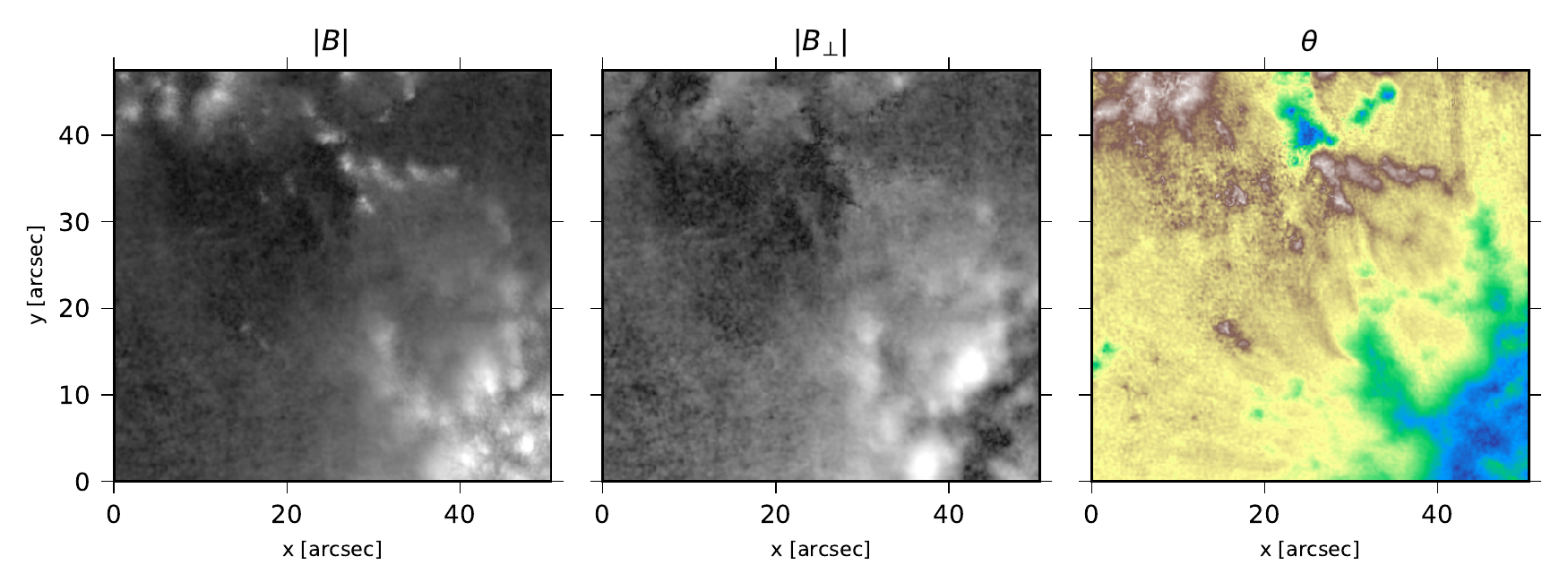}
    \includegraphics[width=0.9\textwidth]{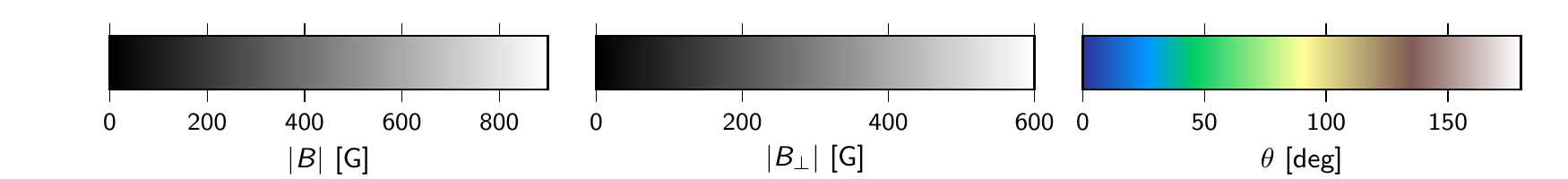}
    \caption{Inferred $|B|$, $|B_\bot|$ and $\theta$ (inclination) maps for spectral region (c) (i.e., based on \ion{Mg}{i}~5173~\AA\ core spectral window).}
    \label{Bhor_azim}
    \end{figure*}
 
 \subsection{Reconstruction of $B_\bot$} \label{sec:result_bhor}
 While we have applied our method to all spectral windows, only the results from spectral window (c) could be meaningfully interpreted as the data in the other spectral windows did not have sufficient signal-to-noise over the entire plage patch. Figure~\ref{Bhor_azim} presents the resulting maps of $|B|$, $|B_\bot|$ and $\theta$. The $|B|$ map confirms the results from our inference of $B_\parallel$ and the entire region seems to be covered by a magnetic canopy with a mean value of 658\,G. From the maps of $B_\bot$ it also becomes clear that the magnetic field in the center of the patch is more vertical and becomes stronger only closer to the edges. The strongest horizontal values that we get are around 607\,G and they are anchored at the edge of the patch with an approximate inclination of $\theta=83^\circ$ relative to the line-of-sight. The mean inclination inside the plage patch is about $25^\circ$ in the observer's reference frame.

\subsection{Reconstruction of the depth stratification}    \label{sec:slice}
    \begin{figure}
    \centering
    \includegraphics[width=\columnwidth, trim=0 0 0 10cm, clip]{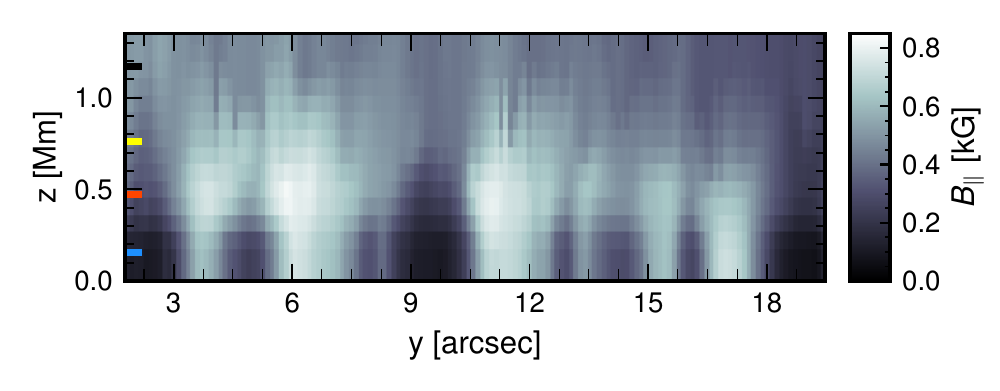}
    \caption{Vertical reconstruction of the canopy magnetic field from the observations. The vertical cut corresponds to the blue slit indicated in panel (a) of Fig.~\ref{Blos_all}. The colored ticks indicate the mean formation region of each spectral region indicated in Table~\ref{table:1}: (a) blue, (b) red, (c) yellow (d) black. The $z$-scale has been estimated for each pixel from its atmosphere derived from an NLTE inversion assuming hydrostatic equilibrium.}
    \label{fig:Blos_slice}
    \end{figure}
   
   Using the formation heights that we calculated in \S\ref{sec:sens} we can reconstruct the stratification of $B_\parallel$ along the slit illustrated in Fig.~\ref{Blos_all}. In each pixel we have four values of the magnetic field and an associated $z$-value. We have interpolated $B_\parallel$ in all pixels along the slit to a equidistant $z$-scale and the resulting stratification is displayed in Fig.~\ref{fig:Blos_slice}. On the left vertical axis we have indicated the mean formation height of each spectral window using colored ticks.
    
    Assuming that our approximations are valid, this figure illustrates that the edge of the magnetic canopy must be located between 300\,km and 600\,km from the continuum formation layer. Table~\ref{table:1} summarizes some statistics of the reconstruction of $B_\parallel$. Although the mean magnetic field value does not change significantly as a function of height from 500\,km to 1000\,km, the large change in the standard deviation illustrates the effect of having a more homogeneous value in the upper layers and more confined and extreme values deeper down. The mean canopy magnetic field value at $z=1000$\,km is $B_\parallel=449$\,G.

\section{Conclusions}
We have implemented a spatially-regularized weak-field approximation by imposing Tikhonov regularization. This type of $\ell$-2 regularization has been commonly used in different stellar Doppler imaging applications \citep{1990MmSAI..61..577P,2002A&A...381..736P,2015ApJ...805..169R,2017A&A...597A..58K}, but to our knowledge this is the first time it is used in combination with the WFA in solar applications with spatially resolved data. 
This method exploits the sparsity of solar data to find solutions that are spatially coherent. 
Our implementation of the spatially-regularized WFA is publicly available\footnote{\url{https://github.com/morosinroberta/spatial_WFA}}.
    
Setting spatial constraints on the WFA reduces the noise present in the reconstructed maps and it improves the fidelity of the reconstruction by coupling the solution spatially. However, the spatial resolution that this method is capable of achieving is still set by the telescope diffraction limit, by the spatial and spectral sampling of the data and by the noise level. The latter is the dominant factor in the upper-right part of our grid (Fig.~\ref{Blos_grid} and \ref{Bhor_grid}) as the higher spatial frequencies of the signal cannot be distinguished from noise any longer and we can only aim at retrieving larger scale details. In our study the parameter dependence is linear and therefore, the regularized WFA method only required a modification of the left-hand side term in the WFA equations. 

The estimation of errors for the WFA becomes less obvious because now the magnetic field solution depends on more than one pixel and it would imply propagating the errors though a system of equations. The formulas provided by \citet{martinez} allow estimating errors based on a less-constrained solution and therefore they can be considered an upper limit of the error estimate.

We have applied the method to high spatial resolution observations of plage acquired in the \ion{Mg}{i}~5173, \ion{Na}{i}~5896 and \ion{Ca}{ii}~8542 \AA\ lines. The WFA assumes a constant stratification of $\boldsymbol{B}$. To overcome this limitation we have defined four narrow spectral windows from observations in three different lines, in order to reconstruct the line-of-sight stratification from the photosphere to the chromosphere. Within each of these narrow spectral windows we have assumed that the amplitude of the Stokes parameters can be described by a constant value of the magnetic field vector. Although a careful application of the methods presented in this paper could allow for the calculation of electrical current densities though Ohm's law, our observations did not achieve the required S/N ratio to get a meaningful stratification of the transverse magnetic field component, which is also required for the calculation of $\boldsymbol{j} = \nabla \times \boldsymbol{B}/\mu_{0}$.

We have focused our analysis on plage regions, where we have studied the expansion of the field as a function of height and the formation of magnetic canopies in the chromosphere. We found that in the photosphere the field is concentrated and there are field-free gaps between magnetic patches. These magnetic concentrations expand horizontally closer to the chromosphere forming a quasi-homogeneous canopy above the photosphere. We found that the plage canopy has a mean total magnetic field of 658\,G at approximately $z=760$\,km above the photosphere (leftmost panel Fig.~\ref{Bhor_azim} and Fig.~\ref{fig:Blos_slice}). In the chromosphere our results suggest that the magnetic field must be strongly vertical. This could also explain the lack of strong Stokes~$Q$ and $U$ signal in the 8542~\AA\ line, in contrast to the 5173 and 5896~\AA\ lines whose cores form closer to the canopy lower boundary where the field is more horizontal (due to the horizontal expansion). In addition, these lines are more sensitive to the presence of magnetic fields in the atmosphere so both effects are likely at play.





\begin{acknowledgements}
This project has received funding from the European Research Council (ERC) under the European Union's Horizon 2020 research and innovation program (SUNMAG, grant agreement 759548). JdlCR is supported by grants from the Swedish Research Council (2015-03994) and the Swedish National Space Agency (128/15). 
GV is supported by a grant from the Swedish Civil Contingencies Agency (MSB). 
The Swedish 1-m Solar Telescope is operated on the island of La Palma by the Institute for Solar Physics of Stockholm University in the Spanish Observatorio del Roque de los Muchachos of the Instituto de Astrof\'isica de Canarias. The Institute for Solar Physics is supported by a grant for research infrastructures of national importance from the Swedish Research Council (registration number 2017-00625).
This research has made use of NASA's Astrophysics Data System Bibliographic Services.
\end{acknowledgements}

%
%

\bibliographystyle{aa} 
\bibliography{references}

\begin{appendix}
\section{Other effects of the regularization terms}\label{ap:1}
     
The \ion{Ca}{ii}~8542~\AA\ line exhibits peculiar profiles in plage targets \citep{2013ApJ...764L..11D}. Due to the stratification of the atmosphere and due to the presence of a hot magnetic canopy, the line source function is very shallow in the chromosphere and the resulting line profile has a flat line core. When applying the standard WFA, the denominator in Eq.~(\ref{eq:wfa_vanilla}) tends to be very small in plage regions. Combined with low S/N data, that equation might combine a numerator and denominator that are both very small and dominated by noise, resulting often in predictions of strong magnetic field. By adding regularization terms, we both construct a non diagonal linear system of equations and add a term to the diagonal (the local contribution). The effect of that term is not only a coupling one, but it also stabilizes the denominator when it is very small, by avoiding a division by zero.

\begin{figure}[!ht]
         \centering
         \includegraphics[width=\hsize]{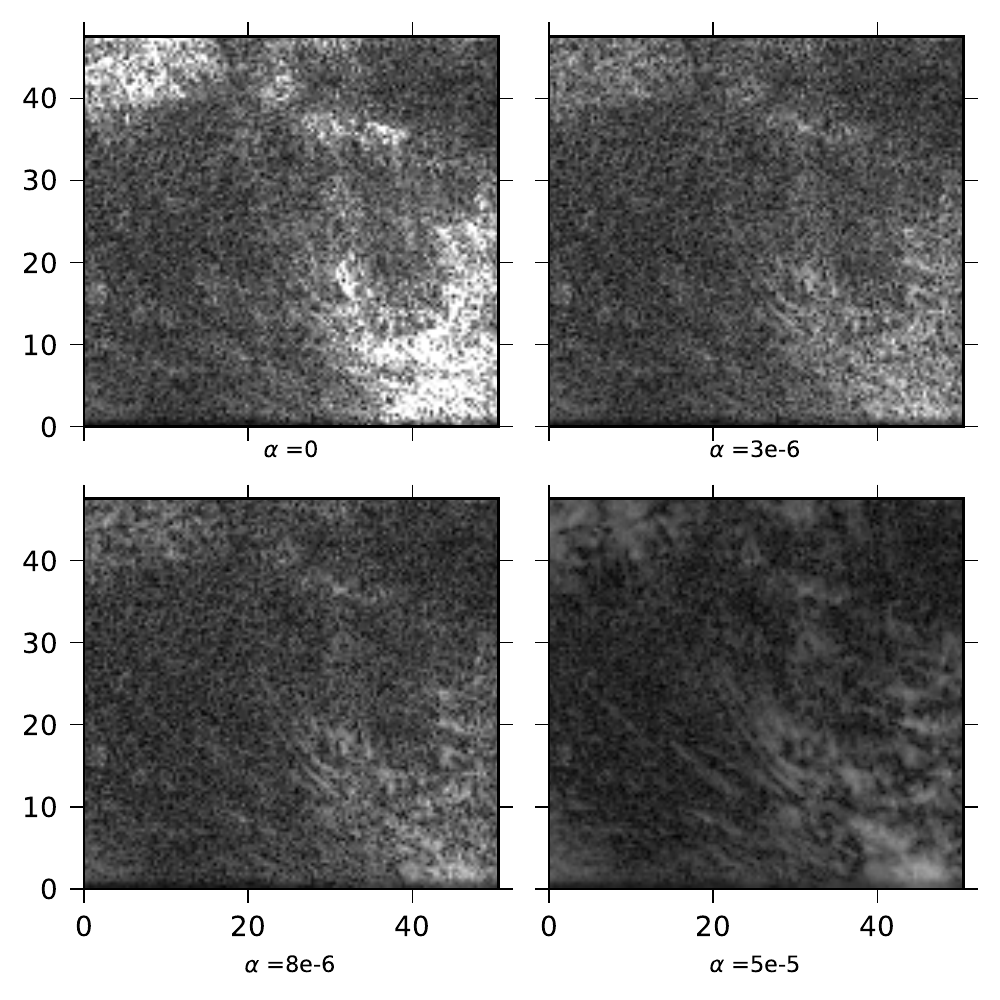}
         \caption{Inferred values of $|B_\bot|$ for the 8542~\AA\ line using an increasing value of he $\alpha$ parameter from left to right. The scaling is set between 0 and 800~G in all panels. The tickmarks are given in arcsec.}\label{fig:Bhorgrid_data}
     \end{figure}

We have illustrated this effect in Fig.~\ref{fig:Bhorgrid_data}, where we have calculated $B_\bot$ using a list of increasing values of the $\alpha$ parameter. In the leftmost panel $\alpha=0$ and the prediction is unrealistically large. We could not see this effect in the simulated data because the simulation does not contain these type of profiles.  We note that even when $\alpha$ is very small and the effect of spatial coupling is barely visible, regularization seems to be enough to remove those unrealistically large predictions ($\alpha=3\times10^{-6}$).

A different type of regularization, which is local, can be applied instead to overcome this particular effect in the WFA. This type of regularization is also an $\ell-2$ method, usually referred to as low-norm regularization. It operates in a similar way as the diagonal damping $\lambda$ parameter that is present in the Levenberg-Marquardt algorithm and it forces the solution to prefer low values of the magnetic field that are compatible with the observations.

The resulting least-squares formula for all components can be expressed as:
\begin{equation}
    B_x = \frac{C\sum_w X_w (I_w')}{C^2\sum_w (I_w')^2 + \beta},\label{eq:beta}
\end{equation}
where $\beta$ is the new regularization parameter, $I_w'$ is the corresponding derivative term for a given line position and $X_w$ is Stokes $Q$, $U$ or $V$. While largely overestimating $\beta$ does not lead to destruction of small scale structures, it suppresses large values of $B_x$, so a careful calibration of the $\beta$ parameter is required when using Eq.~(\ref{eq:beta}).

    
    




\end{appendix}

\end{document}